\documentclass[a4paper,12pt]{article}
\pdfoutput=1

\usepackage{jheppub}
\usepackage[utf8]{inputenc}
\usepackage[english]{babel}

\usepackage{bm}
\usepackage{amsfonts}
\usepackage{amsthm}
\usepackage{mathrsfs}
\usepackage{latexsym}
\usepackage{float}
\usepackage{subfig}
\usepackage[section]{placeins}

\newcommand{\be}{\begin{equation}}
\newcommand{\ee}{\end{equation}}

\newcommand{\benn}{\begin{equation*}}
\newcommand{\eenn}{\end{equation*}}

\usepackage[mathcal]{euscript}
\usepackage{calrsfs}
\DeclareMathAlphabet{\stdcal}{OMS}{zplm}{m}{n}

\newcommand{\pd}{\partial}
\newcommand{\eps}{\varepsilon}
\newcommand{\om}{\omega}
\newcommand{\Ecal}{\stdcal{E}}
\newcommand{\Ocal}{\stdcal{O}}
\newcommand{\Lcal}{\stdcal{L}}
\newcommand{\Odag}{O^\dagger}
\newcommand{\Z}[1]{\mathbb{Z}_{#1}}
\newcommand{\bi}{\mathbf{i}}
\newcommand{\bj}{\mathbf{j}}
\newcommand{\bk}{\mathbf{k}}

\usepackage{mathtools}
\DeclarePairedDelimiter\corrfunc{\langle}{\rangle}
\DeclarePairedDelimiter\bra{\langle}{\rvert}
\DeclarePairedDelimiter\ket{\lvert}{\rangle}
\DeclarePairedDelimiterX\braket[2]{\langle}{\rangle}{#1 \delimsize\vert #2}

\newcommand{\skipline}{\vspace{\baselineskip}}

\newcommand{\figref}[1]{Fig.~\ref{#1}}
\newcommand{\secref}[1]{section~\ref{#1}}

\title{Local quenches in fracton field theory: Lieb-Robinson bound, non-causal dynamics and fractal excitation patterns}

\author[a]{Dmitry S. Ageev,}
\author[b]{Andrey A. Bagrov,}
\author[a]{Aleksandr I. Belokon,}
\author[b]{Askar Iliasov,}
\author[a]{Vasilii V. Pushkarev,}
\author[b]{Femke Verheijen}

\affiliation[a]{Steklov Mathematical Institute of Russian Academy of Sciences,\\
Gubkin str. 8, 119991 Moscow, Russia}
\affiliation[b]{Institute for Molecules and Materials, Radboud University,\\ Heyendaalseweg 135, 6525AJ Nijmegen, The Netherlands}

\emailAdd{ageev@mi-ras.ru}
\emailAdd{andrey.bagrov@ru.nl}
\emailAdd{belokon@mi-ras.ru}
\emailAdd{askar.iliasov@ru.nl}
\emailAdd{pushkarev@mi-ras.ru}

\abstract{We study the out-of-equilibrium dynamics induced by a local perturbation in fracton field theory. For the ${\mathbb Z}_4$ and ${\mathbb Z}_8$-symmetric free fractonic theories, we compute the time dynamics of several observables such as the two-point Green function, $\langle \phi^2 \rangle$ condensate, energy density, and the dipole momentum. The time-dependent considerations highlight that the free fractonic theory breaks causality and exhibits instantaneous signal propagation, even if an additional relativistic term is included to enforce a speed limit in the system. We show that it is related to the fact that the Lieb-Robinson bound does not hold in the continuum limit of the fracton field theory, and the effective bounded speed of light does not emerge. For the theory in finite volume, we show that the fracton wave front acquires fractal shape with non-trivial Hausdorff dimension, and argue that this phenomenon cannot be explained by a simple self-interference effect.}

\makeatletter
\gdef\@fpheader{~}
\makeatother

\begin{document}
\maketitle

\section{Introduction}

Fractons are conjectural states of matter possessing a number of exotic properties that do not allow to describe them within the conventional framework of quantum many-body theory using such concepts as symmetry breaking, band structures, quasiparticles or topological phases~\cite{Pretko:2017fbf, Halasz:2017xoq, Prem:2017kxc, Nandkishore:2018sel, Pretko:2020cko, Gromov:2022cxa}. Continuum fractonic models --- fracton quantum field theories --- have been analyzed in detail~\cite{Pretko:2018jbi, Seiberg:2020bhn, Seiberg:2020wsg, Distler:2021bop} and were shown to be very distinct in many regards from their relativistic counterparts.

Foremost, since the lattice theory spectrum consists of immobile particles or particles with restricted mobility, which are called fractons, at the level of the low-energy theory, there are local defects, whose locations are restricted.

Also, the ground state degeneracy of the system on a lattice scales exponentially with the system size and hence diverges in the continuum limit. Therefore, the continuum version of the theory possesses an infinite number of ground states. In particular, this leads to UV/IR mixing, which means that a low-energy observer is sensitive to short-distance physics. This is a challenge because a continuum field theory is usually expected to give a consistent description of low-energy physics, which does not depend on its UV completion.

Finally, even though fracton field theories are neither Lorentz invariant nor $SO(2)$ rotationally invariant, they possess exotic global symmetries, including subsystem symmetries such that their charges act differently in different sub-spaces of the total space.

Given the very peculiar property of restricted mobility, out-of-equilibrium dynamics of such systems can be highly non-trivial and not resembling that of regular isotropic or smoothly anisotropic theories. By quenching a fracton field theory, one can get a deeper insight into its response properties and unveil features that cannot be deduced from equilibrium considerations. 

In quantum field theory (QFT), quenches have been extensively studied in a number of diverse settings. The number of exactly solvable models is restricted and mainly includes conformal field theories (CFTs)~\cite{Calabrese:2006rx, Calabrese:2007rg, Das:2014jna, Das:2014hqa, Calabrese:2016xau}. Some other results were obtained for free field theories and the simplest interacting theories~\cite{Sotiriadis2009thermal, Sotiriadis:2010si, Rajabpour:2014osa, Das:2015jka, Das:2016lla, MohammadiMozaffar:2018vmk, MohammadiMozaffar:2019gpn, Mozaffar:2021nex, Islam:2023cmm, Ageev:2023wrb}. At the same time, the study of global quench setups is promising as demonstrated by their intensive exploration in terms of AdS/CFT correspondence as a model of thermalization~\cite{Danielsson:1999fa, Balasubramanian:2010ce, Balasubramanian:2011ur, Ageev:2017wet, Ageev:2017oku}.

In~\cite{Ageev:2023wrb}, some of the authors of this paper initiated analysis of time-dependent properties of fractonic matter by considering a global quench in  fracton field theory, where the behavior of the two-point correlation function after the abrupt change of system parameters as well as after an abrupt change of the symmetry of the theory has been analyzed. However, global perturbation of the theory reveals its homogeneous temporal dynamics, but not the spatio-temporal properties, which become visible in the scenario of inhomogeneous --- local --- quench. Hence, in the present paper, we continue this research line and study the dynamics of the  fracton field subject to a local perturbation. Our goal is to see how the structure of fractonic theories affects the propagation of local excitations. In order to contextualize our findings, we put the observables calculated for the fracton field theory in comparison with those in locally quenched relativistic field theory.

In a local quench prescription, the initial state of the system is perturbed locally at one (single quench) or several points (double quench etc.), and its subsequent evolution is reflected in the behavior of quantum mechanical observables --- correlation functions. Local quenches have recently been studied in various theoretical areas, such as condensed matter theory~\cite{Calabrese:2007mtj, cmtloc1, cmtloc2, cmtloc3, cmtloc4, e23020220}, entanglement and quantum gravity~\cite{Calabrese:2005in, Caputa:2014eta, He:2014mwa, Nozaki:2014hna, Cotler:2016acd}, complexity and chaos~\cite{Ageev:2018msv}.

The so-called geometric quench introduced in~\cite{Calabrese:2007rg} involves joining two different theories with a boundary at some time moment. In~\cite{Doyon:2014qsa}, there was made an attempt to generalize this setup to a higher-dimensional case. Another example is the operator local quench, which is a setup describing creating a localized excited state prepared by inserting a local operator into the path-integral that defines the state~\cite{Nozaki:2014hna}. Even though localized excited states have been studied in diverse contexts~\cite{Nozaki:2014hna, He:2014mwa, Astaneh:2014fga, Caputa:2014vaa, Asplund:2014coa, Caputa:2014eta, Guo:2015uwa, Caputa:2015waa, Caputa:2015tua, Chen:2015usa, Ageev:2015qbz, Ageev:2015ykq, Rangamani:2015agy, Caputa:2015qbk, Franchini2015universal, Franchini:2016ivt, He:2017lrg, Guo:2018lqq, Ageev:2018nye, Ageev:2018msv, Shimaji:2018czt, Apolo:2018oqv, Ageev:2018tpd, Caputa:2019avh, He:2019vzf, Bhattacharyya:2019ifi, He:2020qcs, Zenoni:2021iiv, Suzuki:2022xwv}, most of the known examples of local quenches are related to two-dimensional CFT~\cite{Nozaki:2013wia, Caputa:2014vaa, Asplund:2014coa, Caputa:2019avh}. Recently, we have studied operator local quenches in massive scalar field theory~\cite{Ageev:2022kpm}, which is a natural extension of earlier considerations~\cite{Caputa:2014eta}, where the form of observables after a local excitation was completely fixed by conformal symmetry. The method of the straightforward calculation of correlators using Wick's theorem, proposed in this work, allows us to re-derive previously known results and study QFTs with arbitrary interactions without having to confine ourselves to considering only conformal theories. Also, conformal derivation was limited to two dimensions, while this approach works in arbitrary dimensions. The latter fact, in particular, makes it possible to re-derive the results obtained earlier in holographic correspondence AdS$_{d + 1}$/CFT$_d$, where the time evolution of the state on the CFT$_d$ side is dual to a falling massive particle in AdS$_{d + 1}$~\cite{Nozaki:2013wia}.

\skipline

In this paper, we rely on this approach to study operator local quenches in $2 + 1$-dimensional fracton field theory, which is a low-energy limit~\cite{Seiberg:2020bhn} of the XY-plaquette model~\cite{Paramekanti2002ring}. In the latter theory, a compact scalar field, $\phi_s \sim \phi_s + 2\pi$, is defined on each site $s$ of a periodic square lattice with spacing $a$ and the number of sites $L^x$ and $L^y$ in the $x$ and $y$ direction, correspondingly. Its Hamiltonian reads~\cite{Seiberg:2020bhn}
\be
    H = \frac{u}{2}\sum_s \pi^2_s - K\sum_s\cos\left(\Delta_{xy}\phi_s\right), 
    \label{eq:XY-H}
\ee
where $\pi_s$ denotes conjugate momenta of the scalar field, and
\benn
    \Delta_{xy}\phi_{\hat{x},\,\hat{y}} = \phi_{\hat{x} + 1,\,\hat{y} + 1} - \phi_{\hat{x} + 1,\,\hat{y}} - \phi_{\hat{x},\,\hat{y} + 1} + \phi_{\hat{x},\,\hat{y}}.
\eenn
The coordinates $(\hat{x}, \hat{y})$ are integer labels of the sites, $\hat{x} = 1, \ldots, L^x$ and $\hat{y} = 1, \ldots, L^y$. Hamiltonian~\eqref{eq:XY-H} is invariant under momentum dipole symmetries (which are $U(1)_i$ subsystem symmetries along $i$-th coordinate line on the lattice, $\phi_s \to \phi_s + \varphi$) as well as under $\Z4$ rotations of the lattice. The continuum limit of the XY-plaquette model is achieved by taking the double limit $a \to 0, L^i \to \infty$ such that product $a L^i$ is kept fixed. The resulting Lagrangian density is given by~\cite{Seiberg:2020bhn}
\be
    \Lcal = \frac{\mu_0}{2}(\pd_t\phi)^2 - \frac{1}{2\mu}(\pd_x \pd_y \phi)^2, \qquad [\phi] = 0, \quad [\mu_0], [\mu] = 1,
    \label{eq:XY-L}
\ee
where $\mu_0$ and $\mu$ are parameters with mass dimension $+1$. The $U(1)_i$ symmetries of the lattice model become in the continuum limit a momentum dipole symmetry $\phi(t, x,y) \to \phi(t, x,y) + f_x(x) + f_y(y)$ for arbitrary functions $f_x$ and $f_y$. The symmetry under $\Z4$ spatial rotations is preserved in the continuum limit.

\skipline

The paper is organized as follows. In \secref{sec:setup}, we review the operator local quench protocol and the fracton field theory we are studying. The post-quench dynamics of the two-point function is described in \secref{sec:TwoPointFunc}, and the dynamics of the energy and dipole momentum densities --- in \secref{sec:EnergyDensity}. In \secref{sec:LiebRobinson}, we show that the Lieb-Robinson bound of a discretized fractonic theory does not support the emergence of the finite speed of light in the continuous theory. In \secref{sec:FinVol}, we generalize our considerations onto the finite volume case. In \secref{sec:fract_dim}, we comment on the fractal structure of the excitations in the locally quenched theory in finite volume. We also note there that there is a considerable difference between local and global quench setups in fractal behavior of the corresponding excitations. Section \ref{sec:conclusions} is an outlook.

\section{Setup}
\label{sec:setup}
\subsection*{Operator local quench}

An insertion of a local operator~$O$ at spacetime point $(t_0, x_0)$ creates an excited state $\ket{\Psi(t)}$ evolving in time,
\be
    \ket{\Psi(t)} = \stdcal{N}_{O} \cdot e^{-iH(t - t_0)} \cdot e^{-\eps H} O(t_0, x_0)\ket{0},
    \label{eq:operator_insertion}
\ee
where $\stdcal{N}_O$ is a normalization factor, which ensures that $\braket{\Psi}{\Psi} = 1$, and the UV regularization parameter $\eps$ preserves a finite norm of the state. Choosing $(t_0, x_0) = (0, 0)$, the evolution of some observable~$\Ocal$ after the single-point local quench~\eqref{eq:operator_insertion} reduces to a three-point Lorentzian correlation function
\be 
    \corrfunc{\Ocal(t, x)}_{O} = \frac{\bra{0}\Odag(i\eps, 0)\Ocal(t, x)O(-i\eps, 0)\ket{0}}{\bra{0}\Odag(i\eps, 0)O(-i\eps, 0)\ket{0}}.
    \label{eq:our_setting}
\ee
To evaluate it, it is easier to perform calculations in imaginary time and then transform the result to real times by Wick rotation $t \to -i\tau$.\footnote{For a discussion on analytical continuation, we refer the reader to Appendix~A of~\cite{Ageev:2022kpm}.} Usually, the observable~$\Ocal$ is a composite operator, and the calculation requires regularization. One of the possibilities is the point-splitting scheme, where the spacetime point, at which the observable is defined, is split into two separated ones, and then the correlation function~\eqref{eq:our_setting} is calculated using Wick's contractions~\cite{Ageev:2022kpm}. The final correlation function consists of a finite part and a term, which in the limit of merging points contains a constant and a divergence. The sum of the finite part and the constant is interpreted as the actual value of the observable, while the divergent part is discarded (i.e. the correlator is renormalized). For example, in the case of the two-dimensional CFT stress-energy tensor, the constant coming from the point splitting corresponds to the anomalous term in its transformation law~\cite{Ageev:2022kpm}.

At this point, we do not consider theories with interactions, but only free field theories, in which all information is encoded in the two-point correlation functions. In momentum space, it means that the theory is completely determined by its dispersion relation. We will use this machinery to study local quenches in a free QFT with discrete rotational symmetries --- the massive scalar fracton field theory.

\subsection*{Fracton field theory}

In this paper, we consider scalar fracton field theories introduced in \cite{Ageev:2023wrb} with underlying  $\mathbb{Z}_n$ symmetry and a relativistic and a mass terms added for the purpose of regularization. This $\mathbb{Z}_n$  invariance,  translates into the following Euclidean dispersion relation\footnote{Note that in $d = 3$ case, the field $\phi$ has zero mass dimension~\eqref{eq:XY-L}. This means that the coefficient before the $\phi^2$-term has mass dimension $3/2$. We write this coefficient as $\mu m^2$ to deal with the usual dimension-1 mass parameter~$m$, using~$\mu$ to add up one more dimension, so that $[\mu m^2] = 3$.}
\be
    \omega^2 = -\frac{1}{\mu_0}\left[\epsilon\left(k^2 + q^2\right) + \frac{f_{\alpha}(k, q)}{\mu} + \mu m^2\right], \qquad [\epsilon] = 1, \quad [f_\alpha] = 4, \quad \alpha = \frac{n}{4}.
    \label{eq:disp-rel}
\ee
The function $f_{\alpha}(k, q)$ reflects the underlying symmetry of the theory. We also assume that this general form of the dispersion relation includes a rotationally invariant part $\epsilon(k^2 + q^2)$ which tames the divergences corresponding to the UV/IR mixing of fractonic theories, i.e., when $k \to \infty$ and $q \to 0$ or vise versa.

In coordinate space, function $f_\alpha$ corresponds to higher-derivative operators in the action. In the case of $n = 2$, the expression for $f_\alpha$ is given by $k^2q^2$ as follows from~\cite{Seiberg:2020bhn}, see also Appendix~\ref{app:A}. For higher orders, the expression for $f_\alpha(k, q)$ can be derived through the transformation to polar coordinates with stretched angular coordinate, $k = \sqrt{k^2 + q^2}\cos(\alpha\varphi)$ and $q = \sqrt{k^2 + q^2}\sin(\alpha\varphi)$. The result is given by~\cite{Ageev:2023wrb}
\be
    f_\alpha = \left(k^2 + q^2\right)^2\cos^2(\alpha\varphi)\sin^2(\alpha\varphi) = \left[\frac{(k + iq)^{4\alpha} - \left(k^2 + q^2\right)^{2\alpha}}{4\big(k^2 + q^2\big)^{\alpha - 1}\big(k + iq\big)^{2\alpha}}\right]^2.
\ee

Specific examples of such fracton free field theories include:
\begin{itemize}
    \item $\Z4$ fractonic theory, which is invariant under 90$^\circ$ spatial rotations. The Euclidean action is given by the low-energy limit of the XY-plaquette lattice model~\eqref{eq:XY-L} with an added mass term
    \be
        S = \frac{A}{2}\int d\tau\,dx\,dy\left(\mu_0(\partial_{\tau}\phi)^2 + \epsilon(\partial_x\phi)^2 + \epsilon(\partial_y\phi)^2 + \frac{1}{\mu}(\partial_x\partial_y\phi)^2 + \mu m^2\phi^2\right).
    \ee
    The global momentum dipole symmetry of the original theory is now broken by the presence of mass. The dispersion relation is given by~\eqref{eq:disp-rel} with $\alpha = 1$ (see Appendix~\ref{app:A} for details)
    \be
        \omega^2 = -\frac{1}{\mu_0}\left[\epsilon\left(k^2 + q^2\right) + \frac{k^2q^2}{\mu} + \mu m^2\right].
    \ee

    \item $\Z8$ fractonic theory which is invariant under 45$^\circ$ rotations. The dispersion relation is given by~\eqref{eq:disp-rel} with $\alpha = 2$
    \be
        \omega^2 = -\frac{1}{\mu_0}\left[\epsilon\left(k^2 + q^2\right) + \frac{4\big(k^3q - kq^3\big)^2}{\mu\big(k^2 + q^2\big)^2} + \mu m^2\right].
    \ee
\end{itemize}

\section{Dynamics of two-point function and \texorpdfstring{$\phi^2$}{phi2}-condensate}
\label{sec:TwoPointFunc}

In free theory, the dynamics is fully determined by the behavior of two-point correlation functions. Hence two cases with different properties can be studied: 1) when two operators are defined at different spacetime points, and 2) when they are defined at the same point --- so-called $\phi^2$-condensate. In this section, we will consider local quenches corresponding to the insertion of $\phi$ operator.

Let us define equal-time two-point correlation function with the equilibrium part subtracted, namely,
\be
    G(t, x, y) \equiv \frac{\bra{0}\phi(i\eps, 0, 0)\phi(t, x, y)\phi(t, 0, 0)\phi(-i\eps, 0, 0)\ket{0}}{\bra{0}\phi(i\eps, 0, 0)\phi(-i\eps, 0, 0)\ket{0}} - \corrfunc{\phi(t, x, y)\phi(t, 0, 0)}.
    \label{eq:two-point_def}
\ee
A similar one was studied before in the context of global quenches~\cite{Sotiriadis:2010si, Rajabpour:2014osa, Ageev:2023wrb}, and represents a perturbation of the two-point function caused by quench excitation. By the Wick theorem, it transforms to
\be
    \begin{aligned}
        G(t, x, y) & = \frac{\bra{0}\phi(i\eps, 0, 0)\phi(t, x, y)\ket{0}\bra{0}\phi(t, 0, 0)\phi(-i\eps, 0, 0)\ket{0}}{\bra{0}\phi(i\eps, 0, 0)\phi(-i\eps, 0, 0)\ket{0}} + \\ 
        & + \frac{\bra{0}\phi(i\eps, 0, 0)\phi(t, 0, 0)\ket{0}\bra{0}\phi(t, x, y)\phi(-i\eps, 0, 0)\ket{0}}{\bra{0}\phi(i\eps, 0, 0)\phi(-i\eps, 0, 0)\ket{0}}.
    \end{aligned}
\ee
Calculating Wick's contractions using the two-point function in mixed representation~\eqref{eq:mixedprop} we get
\be
    \begin{aligned}
        & \bra{0}\phi(i\eps, 0, 0)\phi(t, x, y)\ket{0}\bra{0}\phi(t, 0, 0)\phi(-i\eps, 0, 0)\ket{0} + \\ 
        & + \bra{0}\phi(i\eps, 0, 0)\phi(t, 0, 0)\ket{0}\bra{0}\phi(t, x, y)\phi(-i\eps, 0, 0)\ket{0} = \\ 
        & = \left(\frac{1}{A}\int\frac{dk\,dq}{(2\pi)^2}\frac{e^{-\om\sqrt{(\eps - it)^2} + ikx + iqy}}{2\om}\right)\left(\frac{1}{A}\int\frac{dk\,dq}{(2\pi)^2}\frac{e^{-\om\sqrt{(\eps + it)^2}}}{2\om}\right) + \text{c.c.},
    \end{aligned}
\ee
with the overall normalization factor given by
\be
    \corrfunc{\phi(i\eps, 0, 0)\phi(-i\eps, 0, 0)} = \frac{1}{A}\int\frac{dk\,dq}{(2\pi)^2}\frac{e^{-2\eps\om}}{2\om}.
\ee

\skipline

\begin{figure}[t]\centering
    \subfloat[relativistic, $t = 1$]{\includegraphics[width=0.33\textwidth]{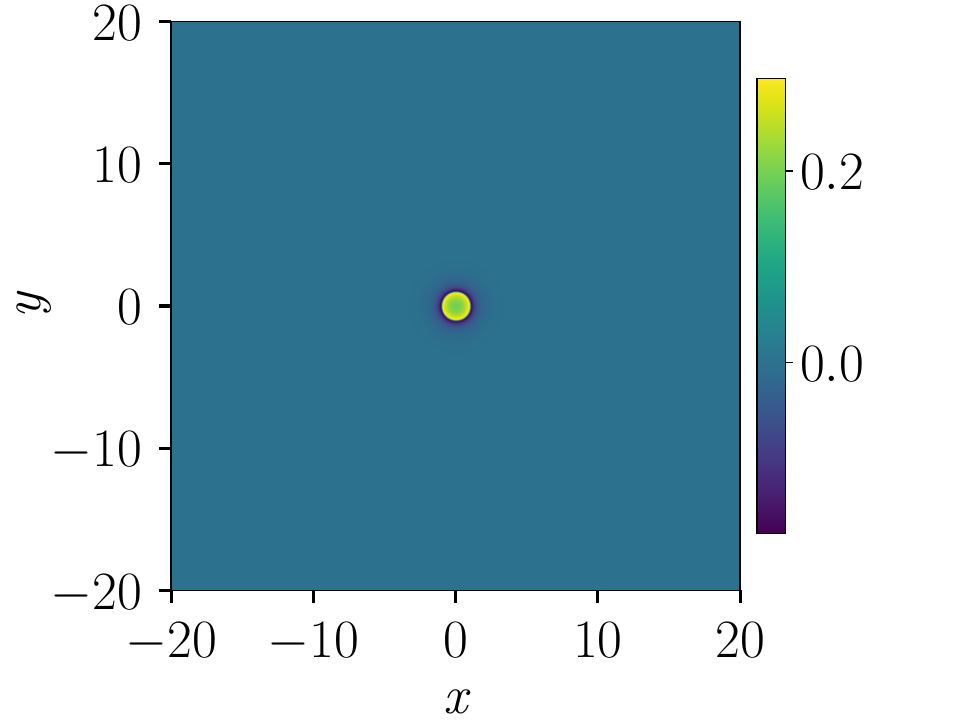}}
    \subfloat[relativistic, $t = 8$]{\includegraphics[width=0.33\textwidth]{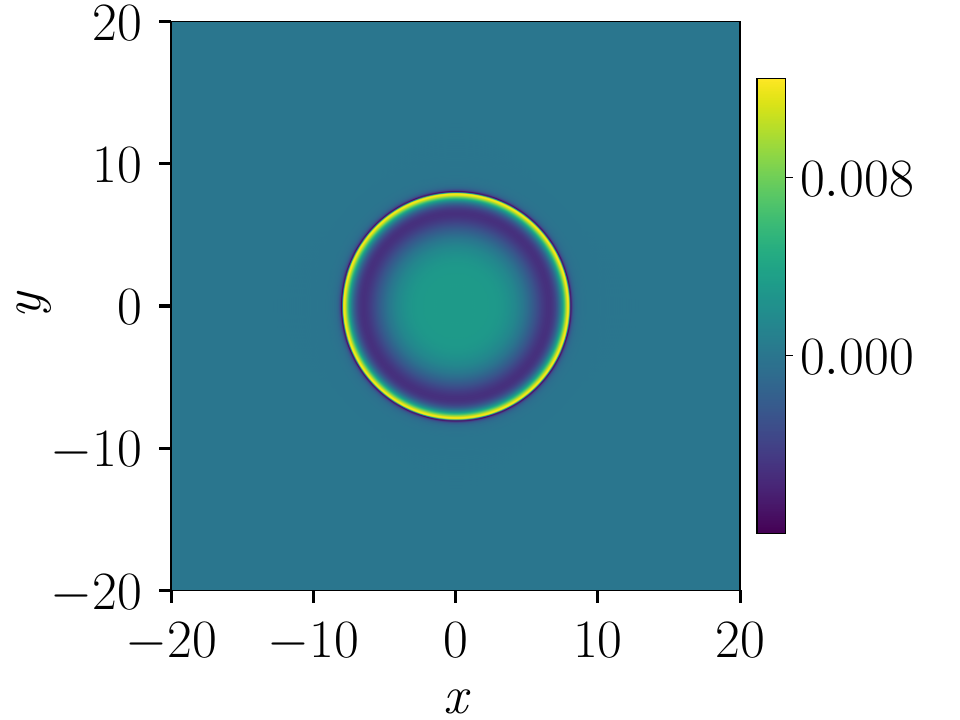}}
    \subfloat[relativistic, $t = 20$]{\includegraphics[width=0.33\textwidth]{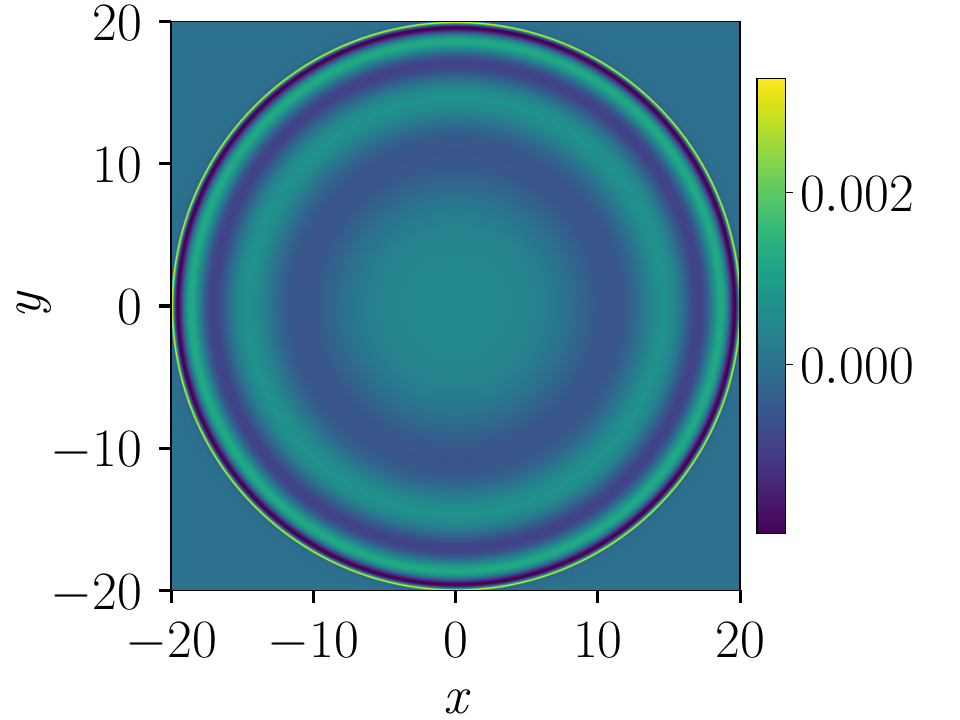}}\\
    \subfloat[$\Z4$, $t = 1$]{\includegraphics[width=0.33\textwidth]{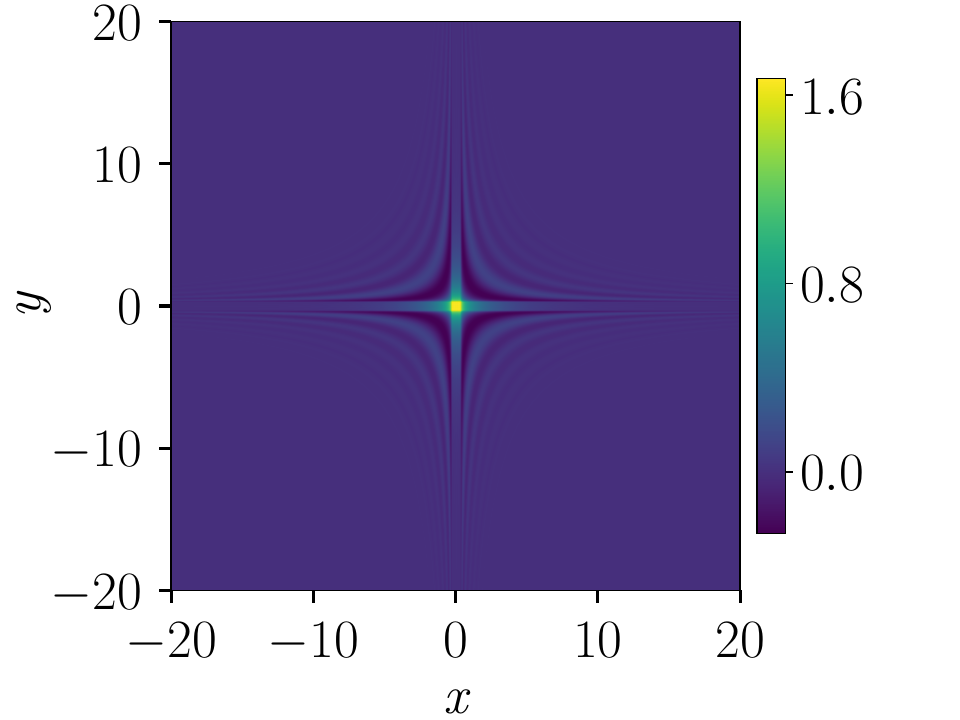}}
    \subfloat[$\Z4$, $t = 8$]{\includegraphics[width=0.33\textwidth]{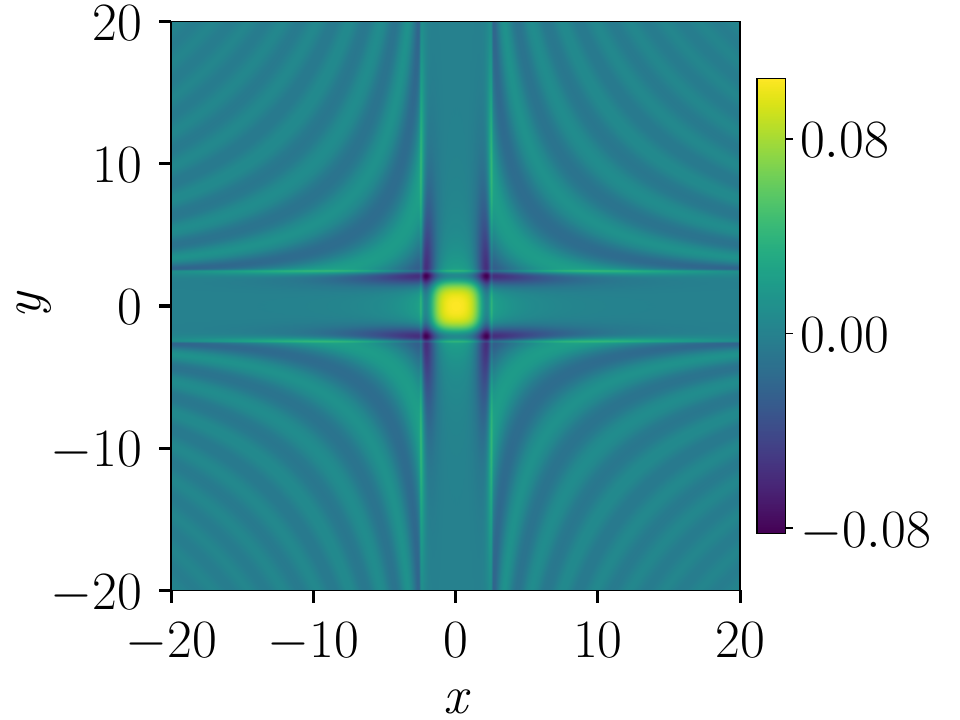}}
    \subfloat[$\Z4$, $t = 20$]{\includegraphics[width=0.33\textwidth]{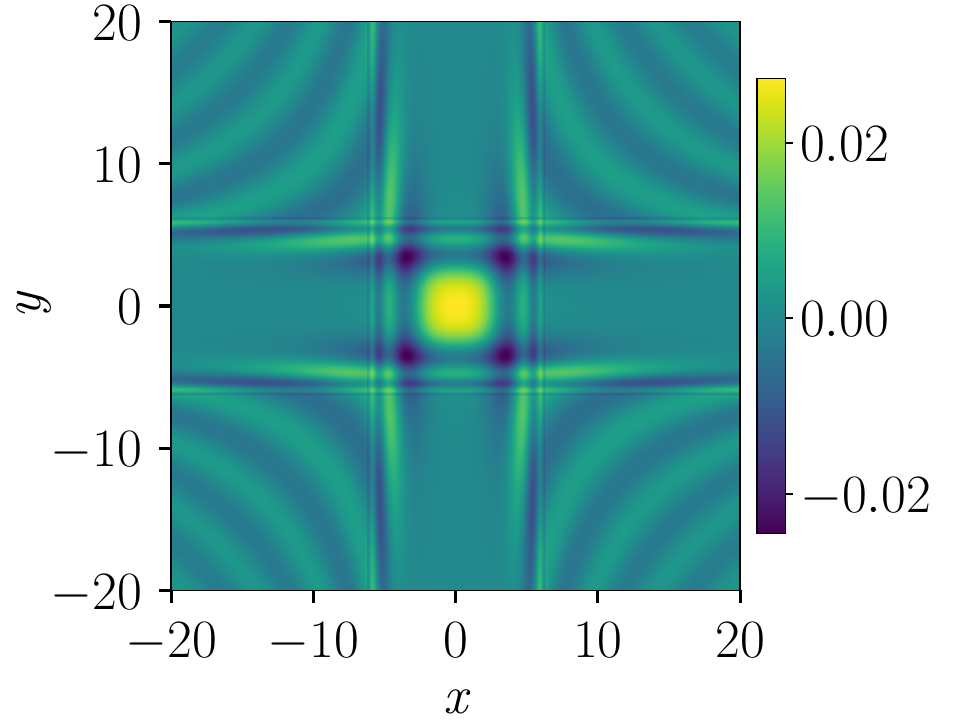}}
    \caption{Dynamics of the non-equilibrium two-point function after a local quench corresponding to operator $\phi$ insertion. \textit{(a) -- (c)}~Spatial profiles of perturbations for fixed time moments in the relativistic model for $m = 1$, $\eps = 0.05$. \textit{(d) -- (f)}~The same for the $\Z4$-symmetric fractonic model; $m = 1$, $\mu_0 = \mu = 1$, $\eps = 0.05$, $\epsilon = 0.1$.}
    \label{fig:2Point}
\end{figure}

The dynamics of the $\phi^2$-condensate after the local quench by operator $\phi$ is given by another type of correlation function, namely,
\be 
    \corrfunc{\phi^2(t, x, y)}_{\phi} = \frac{\bra{0}\phi(i\eps, 0, 0)\phi^2(t, x, y)\phi(-i\eps, 0, 0)\ket{0}}{\bra{0}\phi(i\eps, 0, 0)\phi(-i\eps, 0, 0)\ket{0}}.
    \label{eq:phi2_corr_def}
\ee
Objects of this type have been studied in works on operator local quenches~\cite{Ageev:2022kpm}. To deal with divergences of composite operator $\phi^2$, we split the point at which this operator is inserted into two separate ones, $(t_1, x_1)$ and $(t_2, x_2) = (t_1 + \delta_1, x_1 + \delta_2)$, and take the limit $\delta_{1,2} \to 0$ at the end of the calculation. After that, Wick's contractions are performed, and we obtain
\be
    \begin{aligned}
        & \corrfunc{\phi^2(t, x, y)}_{\phi} = \frac{\bra{0}\phi(i\eps, 0, 0)\phi^2(t, x, y)\phi(-i\eps, 0, 0)\ket{0}}{\bra{0}\phi(i\eps, 0, 0)\phi(-i\eps, 0, 0)\ket{0}} = \\
        & = \lim_{(t_2,\,x_2,\,y_2)\,\to\,(t_1,\,x_1,\,y_1)}\frac{\bra{0}\phi(i\eps, 0, 0)\phi(t_1, x_1, y_1)\phi(t_2, x_2, y_2)\phi(-i\eps, 0, 0)\ket{0}}{\bra{0}\phi(i\eps, 0, 0)\phi(-i\eps, 0, 0)\ket{0}} = \\
        & = 2 \cdot \frac{\bra{0}\phi(i\eps, 0, 0)\phi(t, x, y)\ket{0}\bra{0}\phi(t, x, y)\phi(-i\eps, 0, 0)\ket{0}}{\bra{0}\phi(i\eps, 0, 0)\phi(-i\eps, 0, 0)\ket{0}} + \bra{0}\phi^2(t, x, y)\ket{0},
    \end{aligned}
\ee
where the last term is divergent and should be subtracted. We calculate each contraction numerically starting with the two-point function in mixed representation~\eqref{eq:mixedprop},
\be
    \bra{0}\phi(i\eps, 0, 0)\phi(t, x, y)\ket{0}\bra{0}\phi(t, x, y)\phi(-i\eps, 0, 0)\ket{0} = \left|\frac{1}{A}\int\frac{dk\,dq}{(2\pi)^2}\frac{e^{-\om\sqrt{(\eps - it)^2} + ikx + iqy}}{2\om}\right|^2.
\ee

\begin{figure}[t]\centering
    \subfloat[$\corrfunc{\phi^2}$, $\Z4$, $t = 20$]{\includegraphics[width=0.43\textwidth]{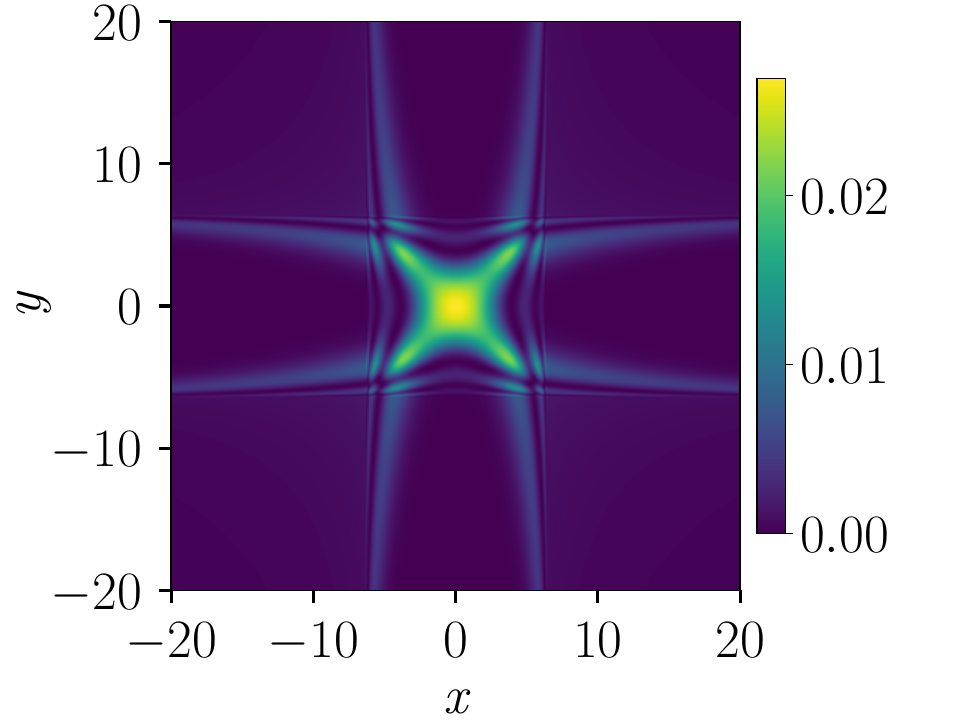}}
    \subfloat[$\corrfunc{\phi^2}$, $\Z4$, $t = 100$]{\includegraphics[width=0.43\textwidth]{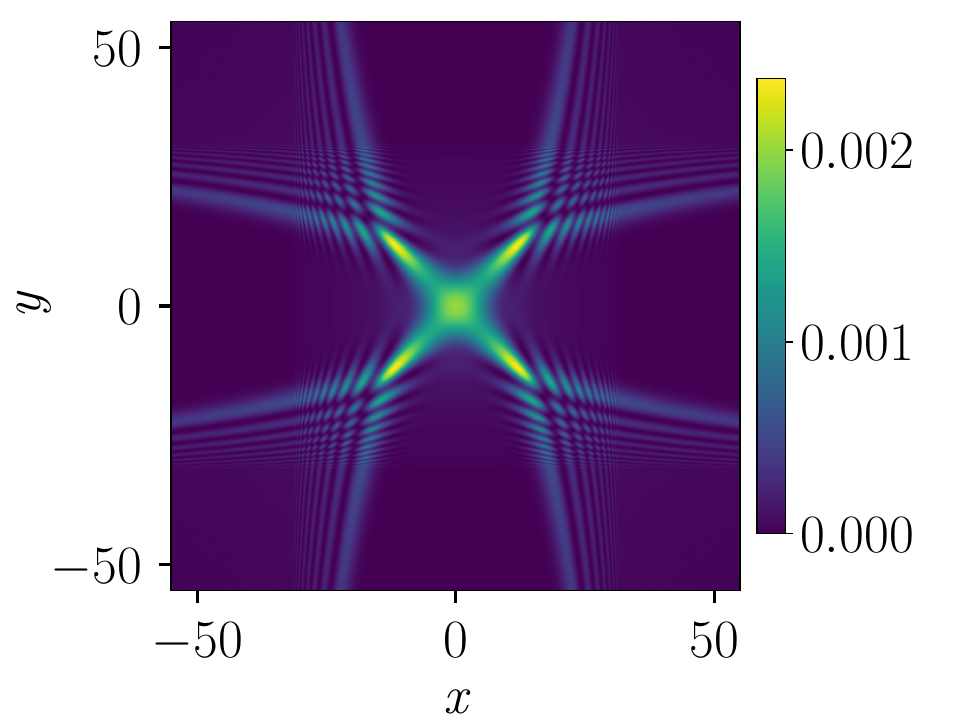}} \\
     \subfloat[$\corrfunc{\phi^2}$, $\Z8$, $t = 20$]{\includegraphics[width=0.43\textwidth]{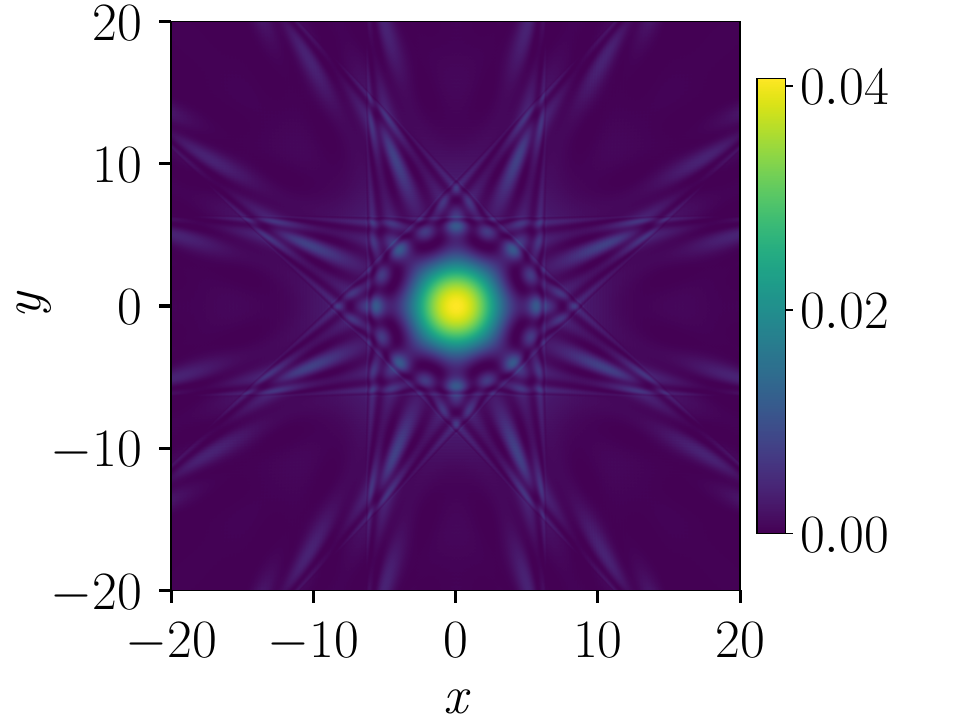}}
     \subfloat[$\corrfunc{\Ecal}$, $\Z4$, $t = 20$]{\includegraphics[width=0.43\textwidth]{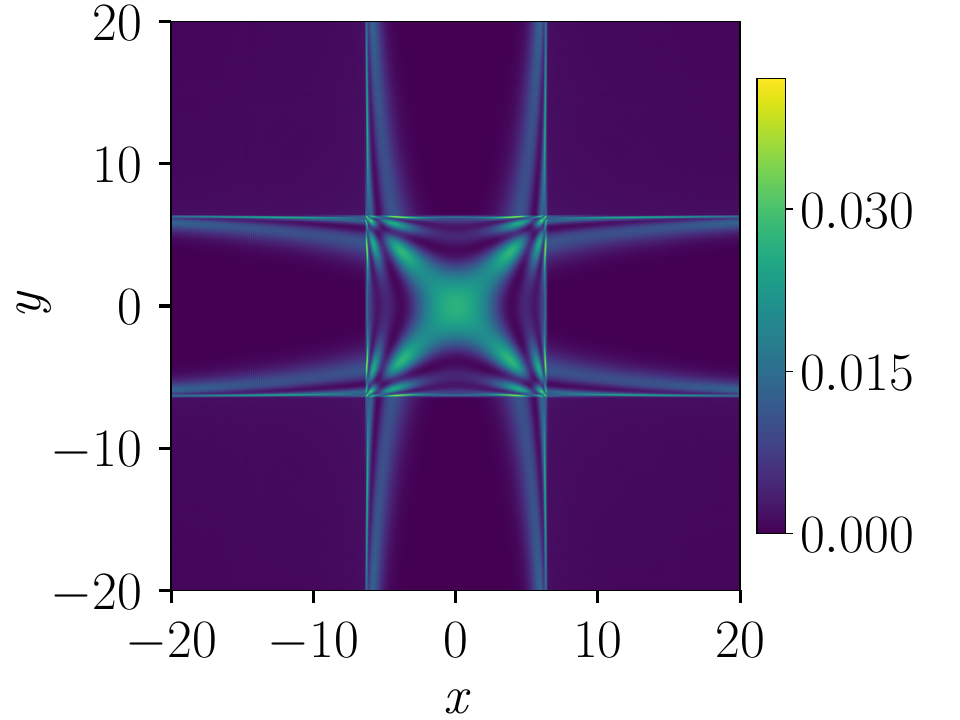}}
    \caption{Dynamics of the observables after a local quench corresponding to insertion of operator $\phi$. \textit{(a), (b)}~Spatial profiles of the perturbations of $\phi^2$-condensate for fixed moments of time $t = 20$ and $t = 100$ in $\Z4$-symmetric fractonic model. \textit{(c)}~Spatial profile of the perturbations of $\phi^2$-condensate for a fixed moment of time $t = 20$ in $\Z8$-symmetric fractonic model. \textit{(d)}~Spatial profile of the perturbations of the energy density for a fixed moment of time $t = 20$ in $\Z4$-symmetric fractonic model. For all cases, $m = 1$, $\eps = 0.05$, $\mu_0 = \mu = 1$, $\epsilon = 0.1$.}
    \label{fig:Phi2_and_EnDens}
\end{figure}

The numerical results for the post-quench dynamics of the relativistic free scalar theory and the fracton field theory are shown in \figref{fig:2Point} for the two-point function, and in \figref{fig:Phi2_and_EnDens} for the $\phi^2$-condensate. One can see that, in the fracton field theory containing the relativistic regulator term, two types of dynamics are present. The severe anisotropy of the fractonic term leads to instantaneous propagation of excitations along the $x-$ and $y-$ axes, and emergence of wave-like patterns beyond the light cone. On the other hand, the relativistic regulator imposes a speed limit on the propagation of square-shaped wave fronts.

\section{Dynamics of the energy density and dipole momentum}
\label{sec:EnergyDensity}

It is instructive to compare the dynamics of the two-point correlation functions with the propagation of energy density~$\Ecal$ given by the following expression,
\be
    \corrfunc{\Ecal(t, x, y)}_{\phi} = \frac{\bra{\phi(i\eps, 0, 0)} \Ecal(t, x, y) \ket{\phi(-i\eps, 0, 0)}}{\corrfunc{\phi(i\eps, 0, 0)\phi(-i\eps, 0, 0)}},
    \label{eq:Edphi}
\ee
where the Euclidean energy density is
\be
    \Ecal = \frac{1}{2}\left(-(\partial_\tau\phi)^2 + (\partial_x\phi)^2 + (\partial_y\phi)^2 + m^2\phi^2\right)
\ee
for the relativistic field theory and
\be
    \Ecal = \frac{1}{2}\left(-\mu_0(\partial_\tau\phi)^2 + \frac{1}{\mu}(\partial_x\partial_y\phi)^2 + \mu m^2\phi^2\right)
\ee
for the massive fracton scalar field theory (see also Appendix~\ref{app:C}). Adding the relativistic regulator~$\epsilon$ to the fractonic theory modifies this expression to
\be
    \Ecal = \frac{1}{2}\left(-\mu_0(\partial_\tau\phi)^2 + \epsilon(\partial_x\phi)^2 + \epsilon(\partial_y\phi)^2 + \frac{1}{\mu}(\partial_x\partial_y\phi)^2 + \mu m^2\phi^2\right).
\ee

The corresponding numerical results for the post-quench dynamics are shown in \figref{fig:Phi2_and_EnDens}.

\begin{figure}[t!]\centering
    \includegraphics[width=0.45\textwidth]{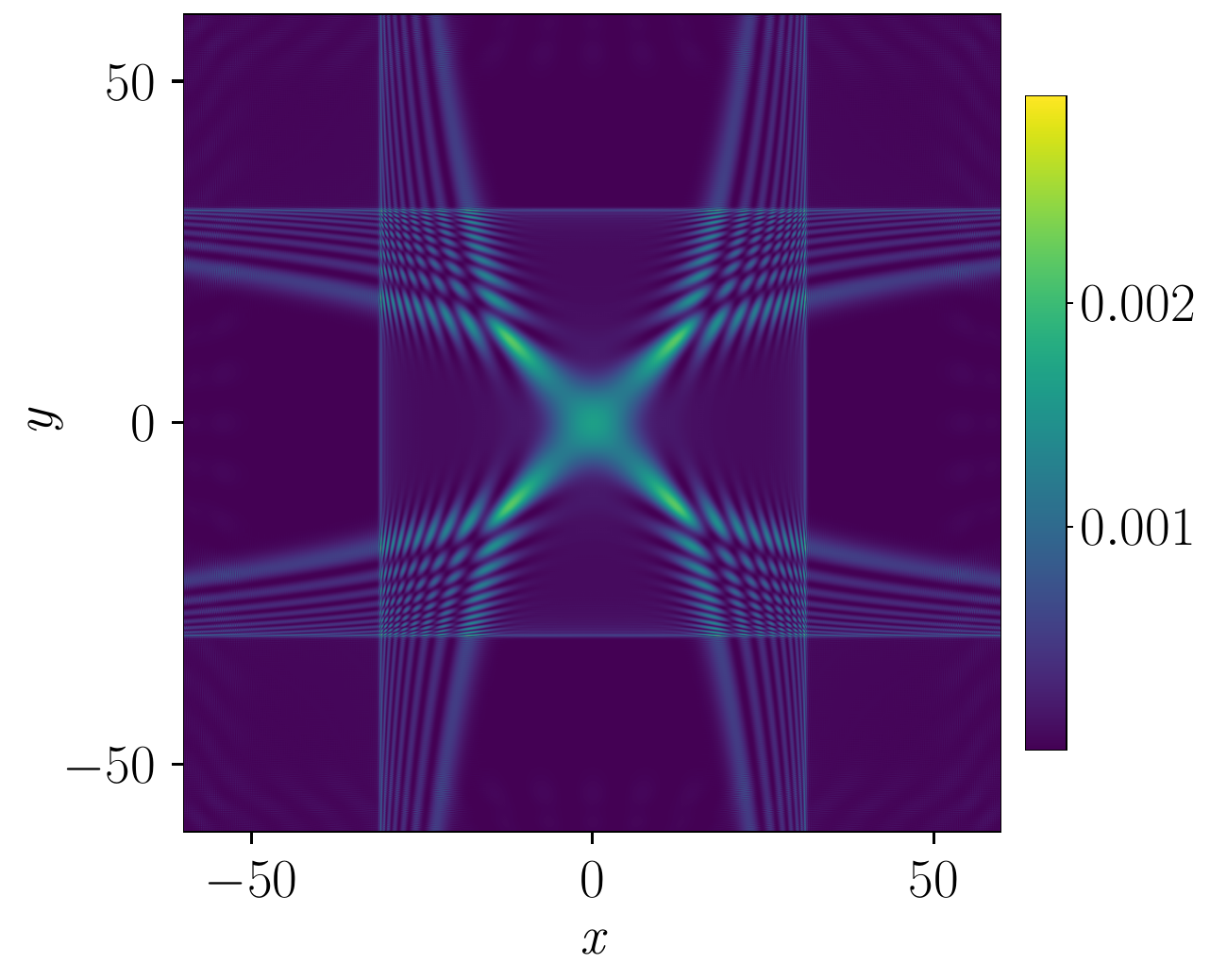}
    \caption{Spatial profile of $\corrfunc{J^2_0}$ after a local quench corresponding to operator $\phi$ insertion for a fixed moment of time $t=100$ for $m = 1$, $\eps = 0.05$, $\mu_0 = \mu = 1$, $\epsilon = 0.1$.}
    \label{fig:dipole_plane}
\end{figure}

Another observable important in this context is the dipole momentum, since its conservation in time is a characteristic feature of fractonic theories. While, strictly speaking, by introducing the relativistic regularization of the UV/IR mixing we violate the dipole momentum conservation, it is still interesting to look at its dynamics. Specifically, we consider the post-quench dynamics of the square of the time component of dipole, given by the following correlation function (see Appendix~\ref{app:C} for details),
\be
    \corrfunc{J^2_0} = \frac{\bra{\phi(i\eps, 0, 0)} J_0^2(t, x, y)\ket{\phi(-i\eps, 0, 0)}}{\corrfunc{\phi(i\eps, 0, 0)\phi(-i\eps, 0, 0)}},
\ee
where the time component of the dipole current is
\be
    J_0 = \mu_0\partial_t \phi.
\ee

In \figref{fig:dipole_plane}, we show spatial profile of $\corrfunc{J^2_0}$ for $t = 100$. One can see that the general structure is similar to other observables, but a manifold of excitation fronts emerges and leads to a complex interference picture near the corners of the square formed by the wave fronts.

\section{Instantaneous signal propagation and the Lieb-Robinson bound}
\label{sec:LiebRobinson}

As evident from the post-quench evolution profiles shown in the previous sections, a local point-like perturbation of fractonic medium instantaneously excites points far away from the perturbation region, violating causality. Adding a relativistic term to the action changes the excitation patterns but does not recover causal dynamics. On the other hand, the fracton field theory can be viewed as a continuum limit of a many-body lattice quantum theory with local interactions, for which the Lieb-Robinson bound setting an effective lightcone can be expected to exist. Hence, it is important to analyze this apparent discrepancy and understand whether the Lieb-Robinson bound induces the finite speed of signal propagation in the continuum theory.

First, we briefly remind the main conditions, which a lattice many-body quantum model should satisfy to ensure the existence of the Lieb-Robinson bound. For that, we shall closely follow~\cite{Cramer:2008}. Let us consider a general harmonic system on lattice~$\mathcal{L}$, defined by a Hamiltonian of the form
\be\label{eq:general_H}
    H_N = \frac{1}{2}\Big[\sum_{\bi,\,\bj \in \mathcal{L}}\hat{p}_{\bi}P_{\bi,\,\bj}\hat{p}_{\bj} + \hat{x}_{\bi}X_{\bi,\,\bj}\hat{x}_{\bj}\Big],
\ee
with a locality constraint: $X_{\bi,\,\bj} = 0$ for $d(\bi,\,\bj) > R $, where $d(\bi,\,\bj)$ is the length of the shortest path on the lattice between sites $\bi$ and $\bj$, and $R$ is a fixed constant. For many systems including the discretization of the fracton field theory considered here, $P_{\bi,\,\bj} = \delta_{\bi,\,\bj}$, so we shall assume this in the subsequent discussion. Then, introducing rescaled time,
\be
    \tau \equiv \sqrt{\|X\|} \, |t|,
\ee
one can claim that the Lieb-Robinson bound is satisfied~\cite{Cramer:2008},
\be
    \sqrt{\|X\|} \, \|[\hat{x}_{\bi}(t), \hat{x}_\bj] \| \le \frac{\sqrt{R}e^{2 d(\bi,\,\bj)\log\left(\frac{eR\tau}{2d(\bi,\,\bj)}\right)}}{\sqrt{d(\bi,\,\bj)}\left(1 - \left(\frac{eR\tau}{2d(\bi,\,\bj)}\right)^2\right)} \,\,\, \mbox{for}\,\, d(\bi,\,\bj) > eR\tau/2.
    \label{eq:LRbound}
\ee
In this case, an effective lightcone arises, which interior is defined as
\be
    d(\bi,\bj) < c|t| \propto eR\tau/2,
\ee
with effective speed of light $c$ given by
\be
    c \propto \frac{1}{2}eR\sqrt{\|X\|}.
    \label{eq:speed}
\ee
To derive the exact emergent speed of light, more assumptions are required, but it is only important for us whether it is bounded from above or not.

Now, let us construct the discretization of the fracton field theory, which is given in the continuum limit by its Lagrangian density,
\be
    \Lcal = \frac{\mu_0}{2}(\pd_t\phi)^2 - \frac{1}{2\mu}(\pd_x\pd_y\phi)^2 - \frac{\epsilon}
    {2}(\pd_x\phi)^2 - \frac{\epsilon}{2}(\pd_y\phi)^2 - \frac{\mu m^2}{2}\phi^2,
\ee
where we included the mass term and the relativistic regulator. The corresponding Hamiltonian density is
\be
    \stdcal{H} = \frac{\mu_0}{2}(\pd_t\phi)^2 + \frac{1}{2\mu}(\pd_x\pd_y\phi)^2 + \frac{\epsilon}{2}(\pd_x\phi)^2 + \frac{\epsilon}{2}(\pd_y\phi)^2 + \frac{\mu m^2}{2}\phi^2.
\ee
The derivatives can be discretized as
\be
    \begin{aligned}\label{eq:mixed_term_discr}
        \pd_x \phi &\rightarrow \frac{1}{1/N}\Big[\phi_{i_x + 1,\,i_y} - \phi_{i_x,\,i_y}\Big], \\
        \pd_y \phi &\rightarrow \frac{1}{1/N}\Big[\phi_{i_x,\,i_y + 1} - \phi_{i_x,\,i_y}\Big], \\ 
        \pd_x \pd_y \phi &\rightarrow \frac{1}{1/N^2}\Big[\phi_{i_x + 1,\,i_y + 1} - \phi_{i_x + 1,\,i_y} - \phi_{i_x,\,i_y + 1} + \phi_{i_x,\,i_y}\Big],
    \end{aligned}
\ee
where we assumed that the system is defined on a square lattice $[0,L]\times[0,L]$, $i_x$ and $i_y$ denote sites, and $N^2$ is the total number of sites. To simplify consideration, we fix $L = 1$. In contrast with~(2.2) in~\cite{Seiberg:2020bhn}, we introduce here lattice constant $a = 1/N$. On the resulting lattice, it is handy to define finite-difference operators,
\be
    \begin{aligned}
        \Delta_x\phi_{i_x,\,i_y} &\equiv \frac{\phi_{i_x + 1,\,i_y} - \phi_{i_x,\,i_y}}{1/N}, \\
        \Delta_y\phi_{i_x,\,i_y} &\equiv \frac{\phi_{i_x,\,i_y + 1} - \phi_{i_x,\,i_y}}{1/N}.
    \end{aligned}
\ee

The corresponding discretized Hamiltonian is
\be
    \begin{aligned}
        H_N & = \frac{1}{2N^2}\sum_{\left\{i_x,\,i_y\right\}\,\in\, \mathcal{L}}\Big(\mu_0\pi^2_{i_x,\,i_y} + \frac{N^4}{\mu}\left[\phi_{i_x + 1,\,i_y + 1} - \phi_{i_x + 1,\,i_y} - \phi_{i_x,\,i_y + 1} + \phi_{i_x,\,i_y}\right]^2 + \\
        & + \epsilon N^2\left[\phi_{i_x + 1,\,i_y} - \phi_{i_x,\,i_y}\right]^2 + \epsilon N^2\left[\phi_{i_x,\,i_y + 1} - \phi_{i_x,\,i_y}\right]^2 + \mu m^2\phi_{i_x,\,i_y}^2\Big),
    \end{aligned}
\ee
where $\pi$ is the canonical momentum operator. After some transformations, it acquires the following form,
\be
    \begin{aligned}
        H_N & = \frac{\mu_0}{2N^2}\sum_{\left\{i_x,\,i_y\right\},\,\left\{j_x,\,j_y\right\}\,\in\, \mathcal{L}}\Big(\pi_{i_x,\,i_y}\delta_{i_x,\,j_x}\delta_{i_y,\,j_y}\pi_{j_x,\,j_y} + \\ 
        & + \frac{N^4}{\mu\mu_0}\Big[\left(4 + \frac{4\mu\epsilon}{N^2} + \frac{\mu^2 m^2}{N^4}\right)\phi_{i_x,\,i_y}\delta_{i_x,\,j_x}\delta_{i_y,\,j_y}\phi_{j_x,\,j_y} - \\
        & - \left(2 + \frac{\mu\epsilon}{N^2}\right)\phi_{i_x,\,i_y}\delta_{i_x,\,j_x}\delta_{i_y,\,j_y}\left(\widetilde\Delta_x + \widetilde\Delta_y\right)\phi_{j_x,\,j_y} + \phi_{i_x,\,i_y}\delta_{i_x,\,j_x}\delta_{i_y,\,j_y} \widetilde\Delta_{x,\,y}\phi_{j_x,\,j_y}\Big]\Big),
    \end{aligned}
\ee
where we defined lattice operators $\widetilde\Delta$ as
\be
    \begin{aligned}
        \widetilde\Delta_x\phi_{i_x,\,i_y} &\equiv \phi_{i_x + 1,\,i_y} + \phi_{i_x - 1,\,i_y}, \\
        \widetilde\Delta_y\phi_{i_x,\,i_y} &\equiv \phi_{i_x,\,i_y + 1} + \phi_{i_x,\,i_y - 1}, \\
        \widetilde\Delta_{x,\,y}\phi_{i_x,\,i_y} &\equiv \phi_{i_x + 1,\,i_y + 1} + \phi_{i_x - 1,\,i_y + 1} + \phi_{i_x + 1,\,i_y - 1} + \phi_{i_x - 1,\,i_y - 1}.
    \end{aligned}
\ee
In the intermediate calculation, we used the translational invariance and reshuffled the indices.

Introducing shorthand notations, $\delta_{\bi,\,\bj} \equiv \delta_{i_x,\,j_x}\delta_{i_y,\,j_y}$, $\phi_\bi \equiv \phi_{i_x,\,i_y}$, and redefining coordinate (field) and momentum operators as
\be
    \begin{aligned}
        \hat{x}_{\bi} &\equiv N^{-1}\phi_{\bi}, \\
        \hat{p}_{\bi} &\equiv N^{-1}\pi_\bi,
    \end{aligned}
\ee
we can rewrite the Hamiltonian in the conventional harmonic form,
\be
    H_N = \frac{\mu_0}{2}\Big[\sum_{\bi,\,\bj\,\in\,\mathcal{L}}\hat{p}_{\bi}P_{\bi,\,\bj}\hat p_{\bj} + \hat{x}_{\bi}X_{\bi,\,\bj}\hat{x}_{\bj}\Big],
\ee
with $P_{\bi,\,\bj} = \delta_{\bi,\,\bj}$ and $X_{\bi,\,\bj} = X\delta_{\bi,\,\bj}$, where
\be
    X = \frac{N^4}{\mu\mu_0}\left[f(N, \mu, \epsilon) - g(N, \mu, \epsilon)\left(\widetilde\Delta_x + \widetilde\Delta_y\right) + \widetilde\Delta_{x,\,y}\right],
\ee
and
\be
    \begin{aligned}
        & f(N, \mu, \epsilon) = 4 + \frac{4\mu\epsilon}{N^2} + \frac{\mu^2 m^2}{N^4}, \\
        & g(N, \mu, \epsilon) = 2 + \frac{\mu\epsilon}{N^2}.
    \end{aligned}
\ee

To get a bound on the norm $\|X\|$, we first obtain spectrum of this matrix. Through a direct substitution, it can be shown that any plane wave is an eigenfunction of $X$,
\be
    \sum\limits_{\bj\,\in\,\mathcal{L}} X_{\bi,\,\bj} \exp\left(\frac{2\pi i}{N}\bk \cdot \bj\right) = \lambda_\bk \exp\left(\frac{2\pi i}{N}\bk \cdot \bi\right),
\ee
where $\bk = (k_x, k_y)$ and $k_x, k_y, i_x, i_y, j_x, j_y$ take integer values from $0$ to $N - 1$, with the eigenvalues
\be\label{eq:lambda_k}
    \lambda_\bk = \frac{16 N^4}{\mu \mu_0}\sin^2\left(\frac{\pi k_x}{N}\right) \sin^2\left(\frac{\pi k_y}{N}\right) + \frac{4 N^2 \epsilon}{\mu_0}\left(\sin^2\left(\frac{\pi k_x}{N}\right) + \sin^2\left(\frac{\pi k_y}{N}\right)\right) + \frac{\mu m^2}{\mu_0}.
\ee 

We shall show that the emergent speed of light in the fractonic theory can be unbounded from above. Let us pick a particular norm, $\|X\| = \|X\|_2$, namely, the spectral norm, in \eqref{eq:speed}. Then, the following identity holds,
\be
    \|X\|_2 = |\lambda|_{\text{max}}.
\ee
Here, the critical difference between the fractonic theory and the relativistic theory becomes clear. First, let us assume that the dispersion relation contains only the relativistic and massive terms, but not the fractonic part, and take for simplicity $\mu = \mu_0 = \epsilon = 1$. Then,
\be
    \|X\|_2 = 8N^2 + m^2,
\ee
and in the continuum limit, one obtains
\be
    c \propto \frac{1}{2}e R\sqrt{\|X\|_2} = \frac{1}{2}\cdot e\cdot \frac{2}{N}\sqrt{8N^2 + m^2} \to 2\sqrt{2} e, \quad \mbox{as} \,\,\, N \to \infty. 
\ee
On the other hand, in the discrete version of the fractonic theory, the leading term in the dispersion relation scales as $\sim N^4$, hence in the continuum limit, $R\sqrt{\|X\|_2} \sim N \to \infty$.

At this point, it is only an indication of the fact that there is no continuum limit of the Lieb-Robinson bound in fracton systems, because in the l.h.s. of \eqref{eq:LRbound}, there is also commutator $\| \left[\hat{x}_\bi(t), \hat{x}_\bj \right] \|$, which can in principle compensate for the rapid growth of $\sqrt{\|X\|}$. Hence, we need to analyze it as well. 

For that, we rely on~(56) from~\cite{Cramer:2008}. If $P_{\bi,\,\bj} \propto \delta_{\bi,\,\bj}$, the commutator of position operators at different times in the harmonic theory can be represented as a series
\be
    C^{xx}_{\bi,\,\bj}(t)\equiv i[\hat x_{\bi}(t), \hat x_{\bj}] = \sum^{\infty}_{n\,=\,0}\frac{(-1)^n t^{2n + 1}(X^n)_{\bi,\,\bj}}{(2n + 1)!},
\ee
that boils down to matrix-valued function
\be
    C^{xx}(t) = \frac{\sin\left(t\sqrt{X}\right)}{\sqrt{X}}.
\ee
It can be evaluated in the diagonal basis of $X$ in the momentum space and then Fourier transformed back to the coordinate representation:
\be
    C^{xx}_{\bi,\,\bj}(t) = \frac{1}{N^2}\sum_{\bk}\frac{\sin(t\sqrt{\lambda_{\bk}})}{\sqrt{\lambda_{\bk}}}\exp\left(\frac{2\pi i}{N}(\bk \cdot \bi - \bk \cdot \bj)\right).
\ee
Prefactor $N^{-2}$ here comes from inversion of the basis vectors.

In the continuum limit, we should rather consider the commutator of field operators
\be
    [\phi_{\bi}(t), \phi_{\bj}] = N^2[\hat{x}_{\bi}(t), \hat{x}_{\bj}],
\ee
which we denote as $\widetilde C^{xx}_{\bi,\,\bj}(t) = N^2 C^{xx}_{\bi,\,\bj}(t)$.

The apparent superluminal signal propagation occurs along the symmetry lines of the fractonic theory, so it is sufficient to analyze behavior of the commutator along one of the coordinate axes. Specifically, we can pick up two points lying on the y-axis, $\bi = \bi_0 = {\bf 0}$ and $\bj = \bj_r = (0, rN)$, with $rN \in \mathbb{Z}$, which turn into $(0, 0)$ and $(0, r)$ in the continuum limit. Then,
\be
    \widetilde C^{xx}_{\bi_0,\,\bj_r}(t) = \sum_{\bk}\frac{\sin(t\sqrt{\lambda_{\bk}})}{\sqrt{\lambda_{\bk}}}e^{-2\pi i r k_y}.
\ee
In the case of the non-regularized fractonic theory with $m = 0$ and $\epsilon = 0$, we can proceed by splitting the sum into three parts: summation over points with momenta $(k_x, 0)$, summation over points with momenta $(0, k_y)$ (for these two groups $\lambda_{\bk} = 0$), and summation over other points ($k_x > 0$, $k_y > 0$). We get
\be\label{eq:comm_sum_split}
    \widetilde C^{xx}_{\bi_0,\,\bj_r}(t) = tN + t\sum^{N - 1}_{k_y\,=\,1}e^{-2\pi i r k_y} + \sum_{k_x, k_y\,=\,1}^{N-1}\frac{\sin(t\sqrt{\lambda_{\bk}})}{\sqrt{\lambda_{\bk}}}e^{-2\pi i r k_y}.
\ee

The next step is to find an upper bound on the last sum in~\eqref{eq:comm_sum_split},
\be
    \left|\sum_{k_x, k_y\,=\,1}^{N - 1}\frac{\sin(t\sqrt{\lambda_{\bk}})}{\sqrt{\lambda_{\bk}}}e^{-2\pi i r k_y}\right| \le \sum_{k_x, k_y\,=\,1}^{N - 1}\frac{\left|\sin(t\sqrt{\lambda_{\bk}})\right|}{\left|\sqrt{\lambda_{\bk}}\right|}\left|e^{-2\pi i r k_y}\right|\le \sum_{k_x, k_y\,=\,1}^{N - 1}\frac{1}{\left|\sqrt{\lambda_{\bk}}\right|}.
\ee
Recalling~\eqref{eq:lambda_k}, the fact that $\sin x >x/2$ for $x\in(0,\pi/2)$, and the symmetry $\sin \alpha = \sin(\pi - \alpha)$, one can obtain the following estimate,
\be
    \sum_{k_x, k_y\,=\,1}^{N - 1}\frac{1}{\left|\sqrt{\lambda_{{\bf k}}}\right|} \le \frac{\sqrt{\mu\mu_0}}{\pi^2}\sum_{k_x, k_y\,=\,1}^{\lfloor\frac{N}{2}\rfloor}\frac{1}{k_x k_y},
\ee
where $\lfloor\frac{N}{2}\rfloor$ denotes the integer part of the fraction. The resulting sum can be estimated from above by integral
\begin{gather}
    \sum_{k_x, k_y\,=\,1}^{\lfloor\frac{N}{2}\rfloor}\frac{1}{k_x k_y} = \left(\sum^{\lfloor\frac{N}{2}\rfloor}_{k\,=\,1}\frac{1}{k}\right)^2\le\left(1 + \int^{\lfloor\frac{N}{2}\rfloor + 1}_{1}\frac{du}{u}\right)^2 = \left(1+\ln\left(\lfloor\frac{N}{2}\rfloor + 1\right)\right)^2.
\end{gather}

The second term of~\eqref{eq:comm_sum_split} is 
\be
    t\sum^{N - 1}_{k_y\,=\,1}e^{-2\pi i r k_y} = t \left(- 1 + \frac{e^{-2\pi irN} - 1}{e^{-2\pi ir} - 1}\right),
\ee
and for any $r\in(0,1)$,
\be 
    \left|\frac{e^{-2\pi irN} - 1}{e^{-2\pi ir} - 1}\right| \leq \frac{2}{\left| e^{-2\pi ir} - 1\right|}.
\ee

Now, using the fact that, if $a = b + c + d$ and $|c| \le f$, $|d|\le g$, then $a\ge b - f - g$, one can derive the following lower bound on the commutator,
\be
    \left|\widetilde C^{xx}_{\bi_0,\,\bj_r}(t)\right| \ge t(N - 1) - \frac{2t}{\left|e^{-2\pi ir} - 1\right|} - \frac{\sqrt{\mu\mu_0}}{\pi^2}\left(1 + \ln\left(\lfloor\frac{N}{2}\rfloor + 1\right)\right)^2.
\ee
Since the linear term grows faster than the logarithmic one, the norm of the commutator is divergent for large $N$, for any distances along the cross $x = 0$ and $y = 0$ and any values of time except $t = 0$. Hence, l.h.s. of \eqref{eq:LRbound}, which in the continuum case takes the form $\sqrt{\|X\|} \, \|[\phi_\bi(t), \phi_\bj]\|$, is a product of two terms unbounded from above, and the Lieb-Robinson bound is satisfied only for smaller and smaller values of times $t$ as the lattice spacing goes to zero. As a result, in the continuum limit, the effective speed of light in the fractonic theory can be unbounded.

This calculation also explains why introducing the massive term and the relativistic regulator does not remove the non-causal signal propagation. In the expression for eigenvalues \eqref{eq:lambda_k}, these terms scale slower with $N$ than the fractonic term, hence cannot compensate for the unlimited growth of $\|X\|$ and $\|[\phi_\bi(t), \phi_\bj]\|$ in the continuum limit.

\section{Local quench in finite volume}
\label{sec:FinVol}

The norm of a state locally perturbed by an operator in a theory defined on a plane is infinite. Hence, it is instructive to study post-quench dynamics in a regularized theory, where the norm divergence is removed by putting the system on a finite torus. Here, we consider the evolution of observables after a local quench with operator $\phi$ on a $2 + 1$-dimensional Lorentzian torus, i.e., $x \sim x + L_x$, $y \sim y + L_y$. Starting with the general form of the two-point function~\eqref{eq:2point_fin_vol}, non-equilibrium Green function~$G(t, x, y)$~\eqref{eq:two-point_def} is calculated via the following Wick's contractions
\be
    \begin{aligned}
        & \corrfunc{\phi(i\eps, 0, 0)\phi(t, x, y)}\corrfunc{\phi(t, 0, 0)\phi(-i\eps, 0, 0)} + \\
        & + \corrfunc{\phi(i\eps, 0, 0)\phi(t, 0, 0)}\corrfunc{\phi(t, x, y)\phi(-i\eps, 0, 0)} = \\
        & = \left(\frac{1}{A L_x L_y}\cdot\frac{e^{-m\sqrt{(\eps - it)^2}}}{2m} + \frac{1}{A L_x L_y}\sum\limits_{n,s\,\neq\,0} e^{i k_n x + i q_s y}\left(\frac{e^{-\om_{ns}\sqrt{(\eps - it)^2}}}{2\om_{ns}}\right)\right) \times \\
        & \times \left(\frac{1}{A L_x L_y}\cdot\frac{e^{-m\sqrt{(\eps + it)^2}}}{2m} + \frac{1}{A L_x L_y}\sum\limits_{n,s\,\neq\,0}\frac{e^{-\om_{ns}\sqrt{(\eps + it)^2}}}{2\om_{ns}}\right) + \text{c.c.},
    \end{aligned}
\ee
and the correlation function of $\phi^2$-condensate $\corrfunc{\phi^2}_\phi$~\eqref{eq:phi2_corr_def},
\be
    \begin{aligned}
        & \corrfunc{\phi(i\eps, 0, 0)\phi(t, x, y)}\corrfunc{\phi(t, x, y)\phi(-i\eps, 0, 0)} = \\
        & = \left|\frac{1}{A L_x L_y}\cdot\frac{e^{-m\sqrt{(\eps - it)^2}}}{2m} + \frac{1}{A L_x L_y}\sum\limits_{n,s\,\neq\,0} e^{i k_n x + i q_s y}\left(\frac{e^{-\om_{ns}\sqrt{(\eps - it)^2}}}{2\om_{ns}}\right)\right|^2
    \end{aligned}
\ee
with the normalization factor
\be
    N = \corrfunc{\phi(i\eps, 0, 0)\phi(-i\eps, 0, 0)} = \frac{1}{A L_x L_y}\cdot\frac{e^{-2m\eps}}{2m} + \frac{1}{A L_x L_y}\sum\limits_{n,s\,\neq\,0}\frac{e^{-2\om_{ns}\eps}}{2\om_{ns}}.
\ee
The sums over Fourier modes will be calculated numerically.

\begin{figure}[ht]\centering
    \subfloat[relativistic, $t = 10$]{\includegraphics[width=0.33\textwidth]{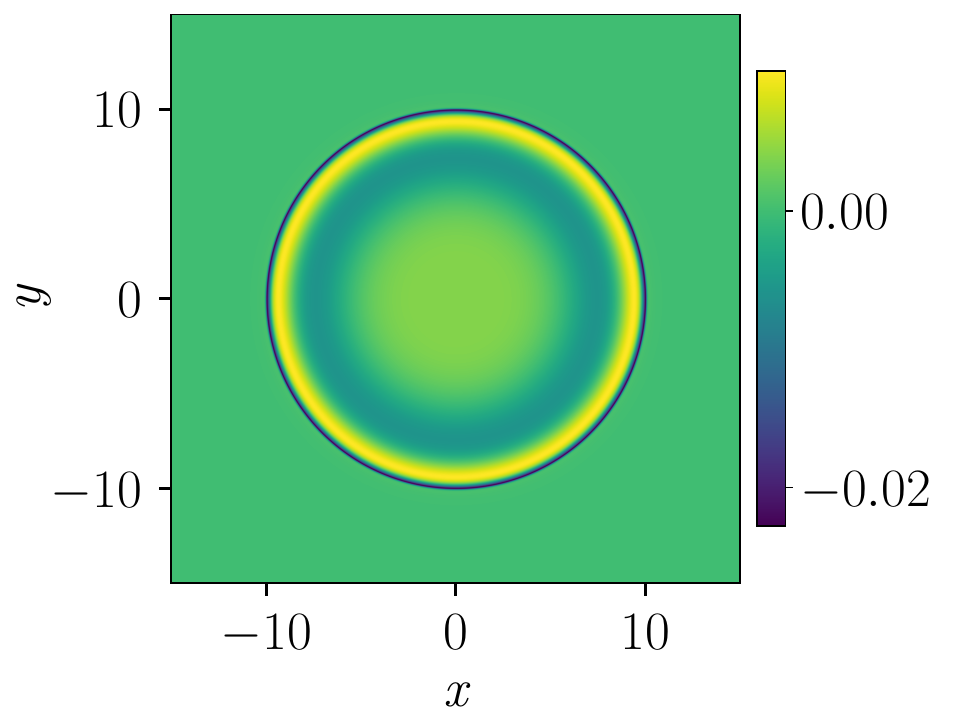}}
    \subfloat[relativistic, $t = 20$]{\includegraphics[width=0.33\textwidth]{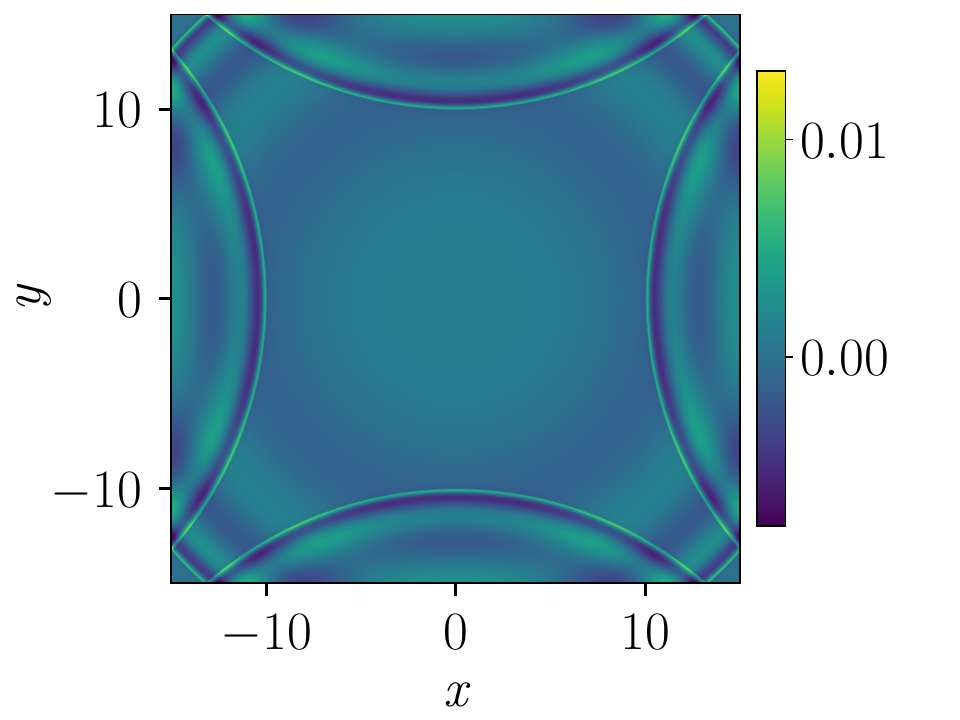}}
    \subfloat[relativistic, $t = 100$]{\includegraphics[width=0.33\textwidth]{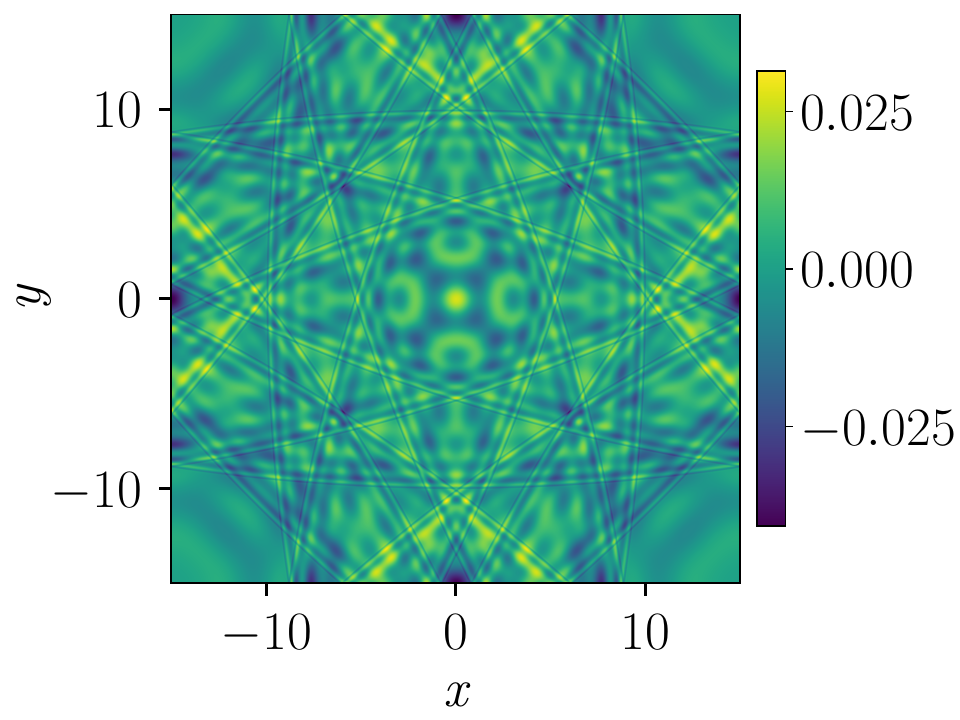}}\\
    \subfloat[$\Z4$, $t = 10$]{\includegraphics[width=0.33\textwidth]{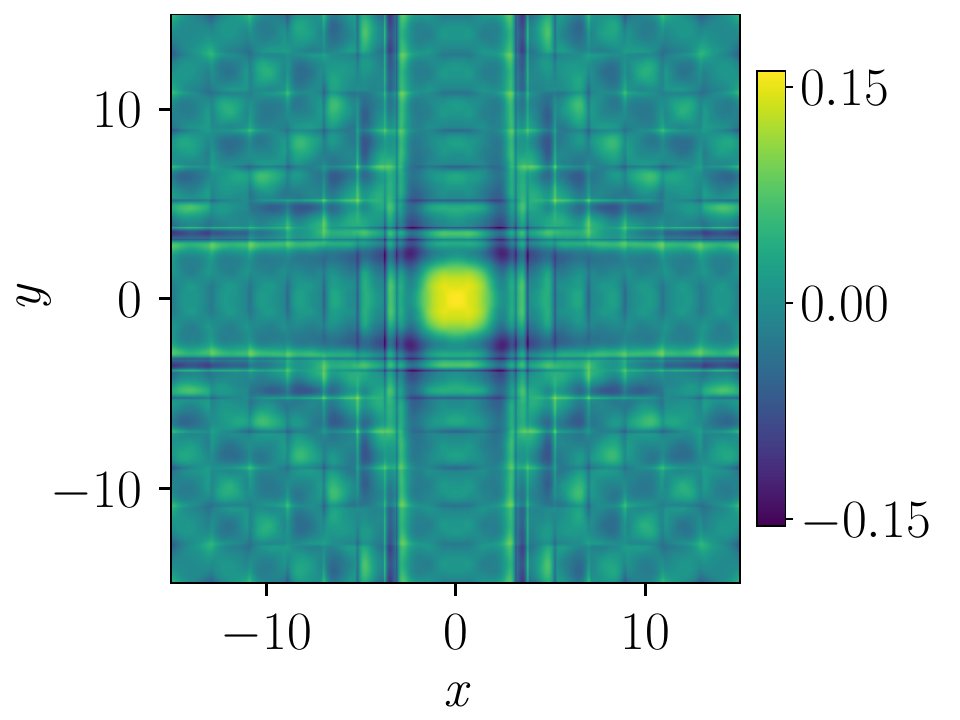}}
    \subfloat[$\Z4$, $t = 20$]{\includegraphics[width=0.33\textwidth]{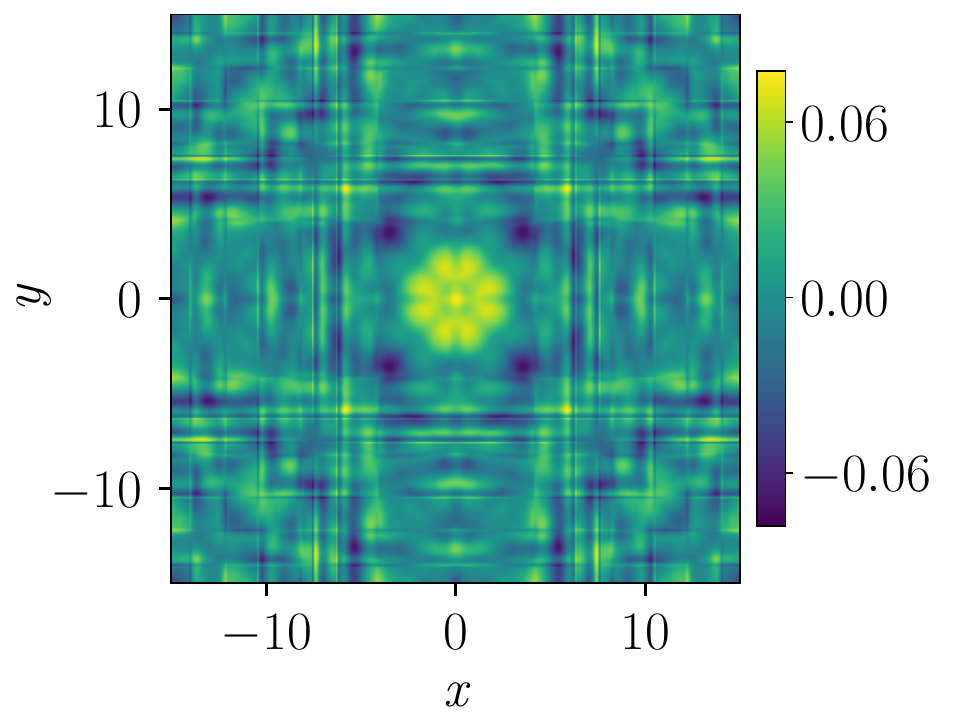}}
    \subfloat[$\Z4$, $t = 100$]{\includegraphics[width=0.33\textwidth]{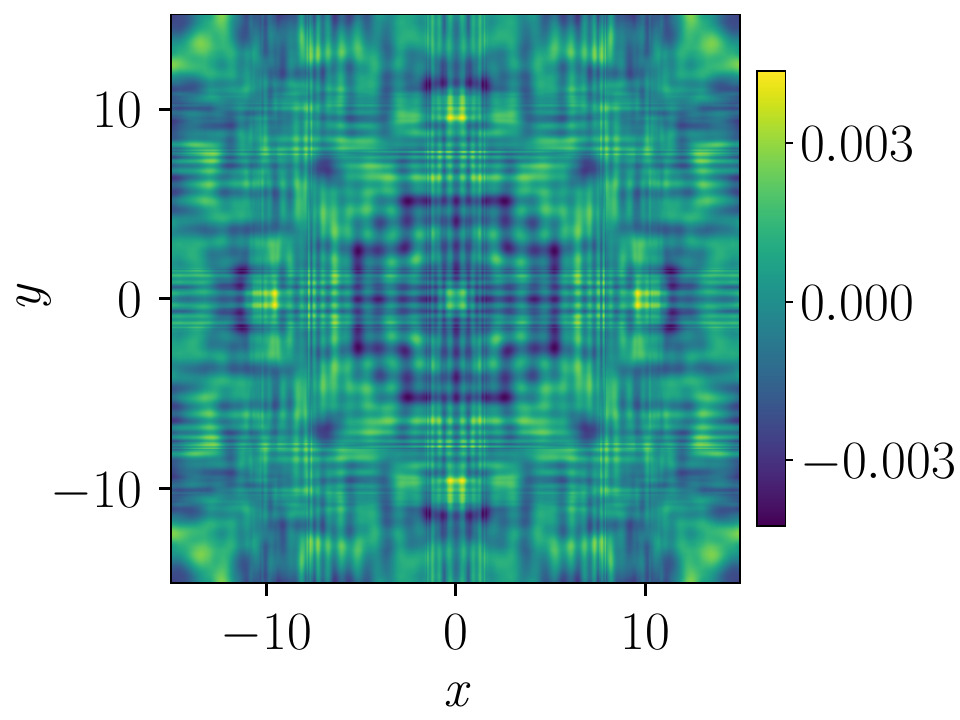}}\\
    \subfloat[$\Z8$, $t = 10$]{\includegraphics[width=0.33\textwidth]{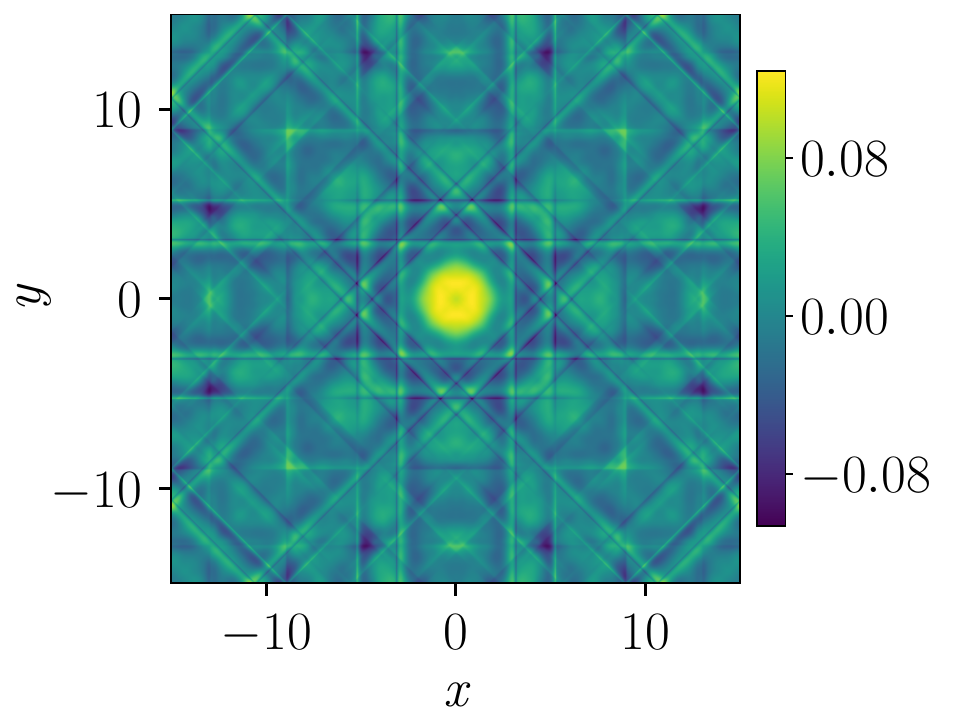}}
    \subfloat[$\Z8$, $t = 20$]{\includegraphics[width=0.33\textwidth]{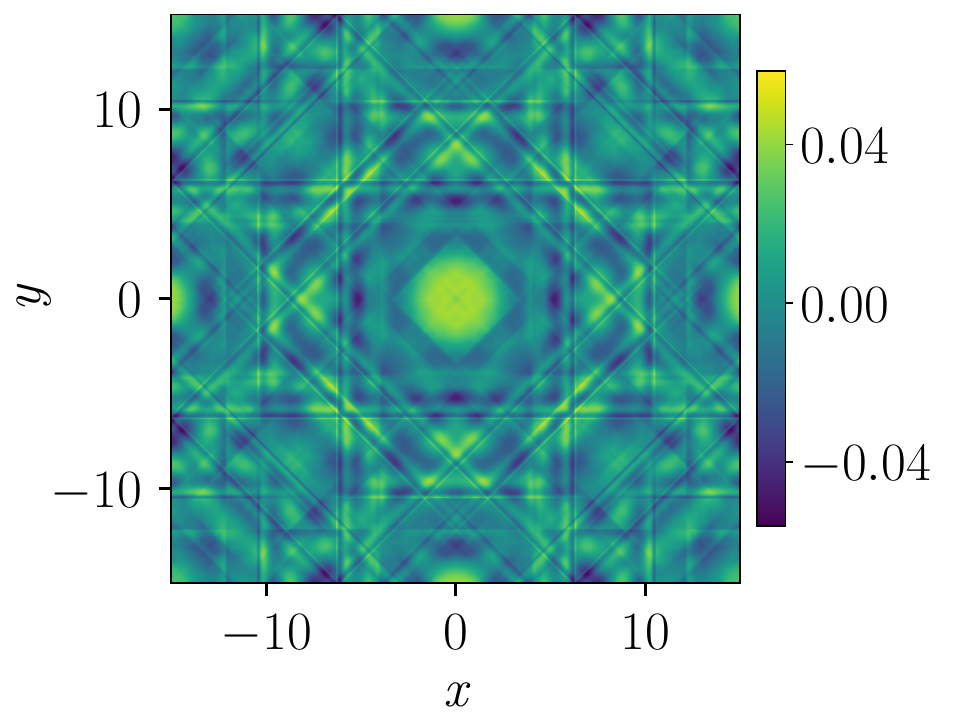}}
    \subfloat[$\Z8$, $t = 100$]{\includegraphics[width=0.33\textwidth]{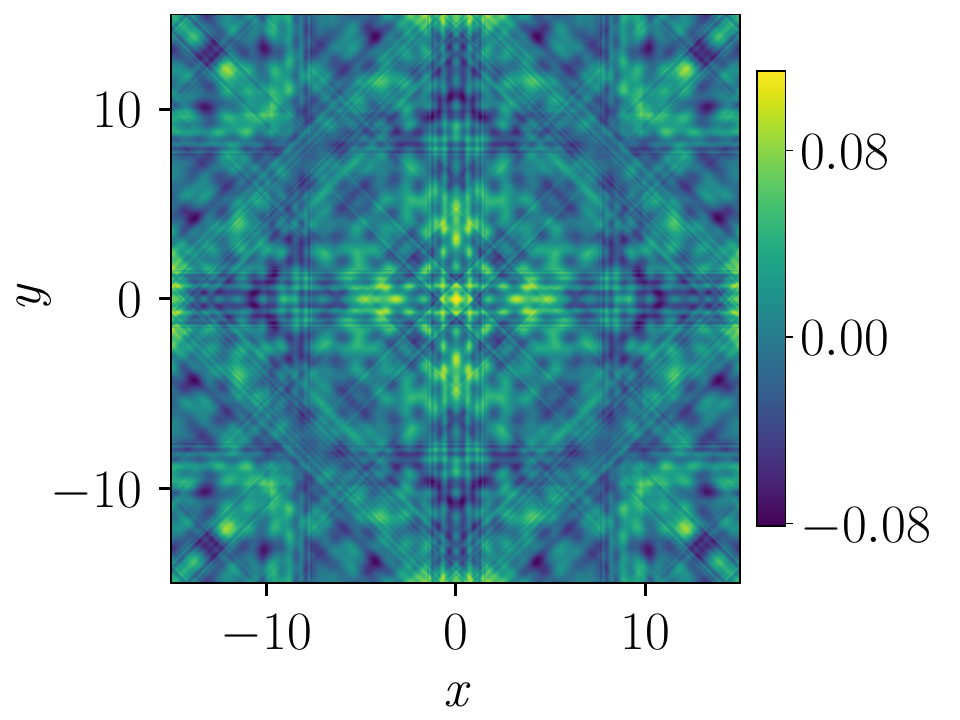}}
    \caption{Dynamics of the non-equilibrium two-point function after a local quench corresponding to the insertion of operator $\phi$. \textit{(a) -- (c)}~Spatial distribution of perturbations for fixed moments of time in the relativistic model. The parameters are: $L_x = L_y = 30$, $m = 1$, $\eps = 0.05$. \textit{(d) -- (f)}~The same for the $\Z4$-symmetric fractonic model; $L_x = L_y = 30$, $m = 1$, $\eps = 0.05$, $\mu_0 = \mu = 1$, $\epsilon = 0.1$. \textit{(g) -- (i)}~The same for the $\Z8$-symmetric fractonic model; $L_x = L_y = 30$, $m = 1$, $\eps = 0.05$, $\mu_0 = \mu = 1$, $\epsilon = 0.1$.}
    \label{fig:2PointFinVol}
\end{figure}

As a warm-up, we first consider the evolution of $\phi^2$-condensate in the Klein-Gordon massive free theory. Note that the massless limit $m \to 0$ of the two-point function~\eqref{eq:2point_fin_vol} does not exist. This is a general result for any number of compact dimensions since the divergent term $e^{-m\sqrt{\tau^2}}/m$ in the series does not depend on $d$. However, in the flat-space limit $L_{x,\,y} \to \infty$, this divergence vanishes, and the two-point function becomes well-defined. Both massless and massive flat-space limits $L_{x,\,y} \to \infty$ can be obtained analytically. For $d = 2 + 1$, we showed in~\cite{Ageev:2022kpm} that
\be
    \corrfunc{\phi^2(t, x, y)}_{\phi,\,3d} \underset{L_x, L_y\,\to\,\infty}{\approx} \frac{4\eps e^{2\eps m}\left|e^{-m\sqrt{(\eps - it)^2 + \rho^2}}\right|^2}{\left|\sqrt{(\eps - it)^2 + \rho^2}\right|^2}, \quad \text{where} \quad \rho^2 = x^2 + y^2.
    \label{eq:d3-phi2}
\ee
In contrast with the local quench in CFT$_2$~\cite{Caputa:2014eta}, after which the perturbation propagates as a soliton-like wave and does not dissipate, in $d = 2 + 1$ dimensions ($d > 2$ in general, see~\cite{Ageev:2022kpm}), the amplitude of the perturbation~\eqref{eq:d3-phi2} decays even in the massless limit.

\begin{figure}[ht]\centering
    \subfloat[$\Z4$, $t = 10$]{\includegraphics[width=0.33\textwidth]{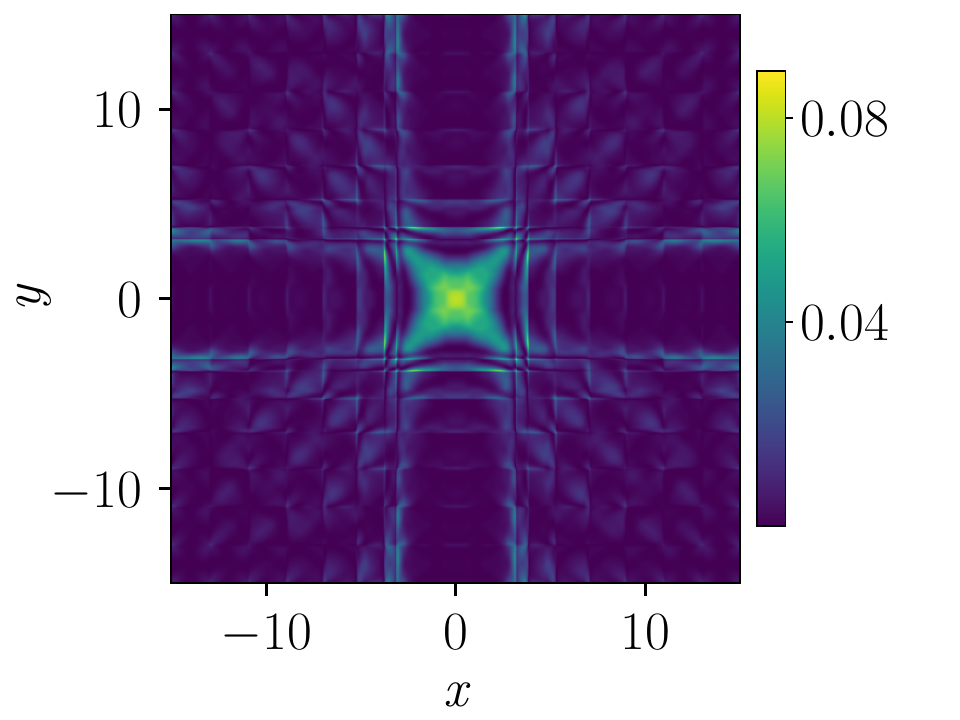}}
    \subfloat[$\Z4$, $t = 20$]{\includegraphics[width=0.33\textwidth]{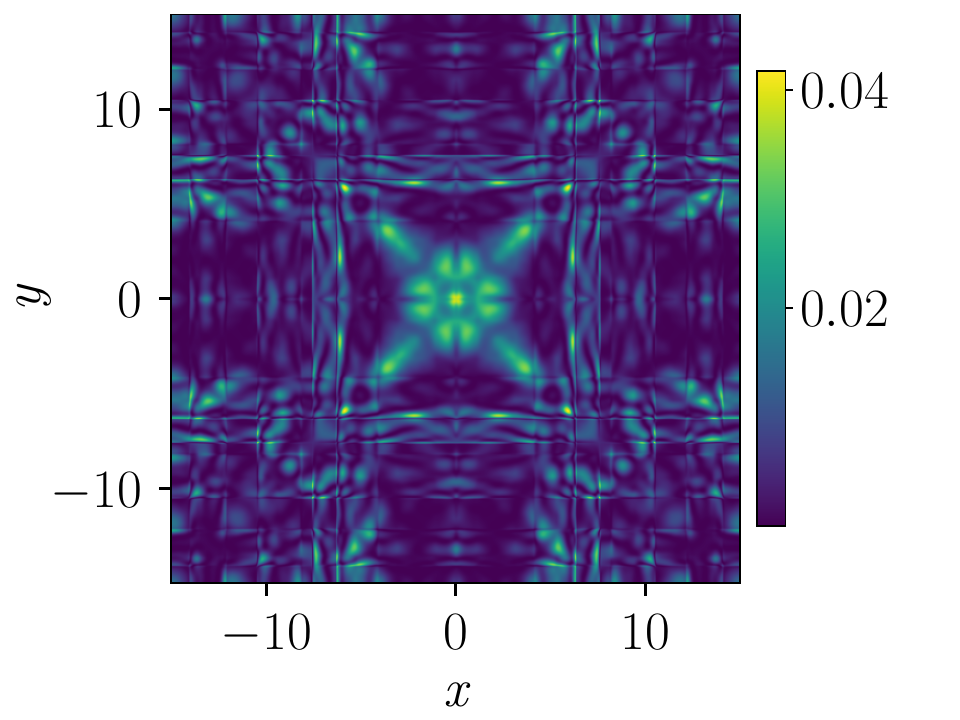}}
    \subfloat[$\Z4$, $t = 100$]{\includegraphics[width=0.33\textwidth]{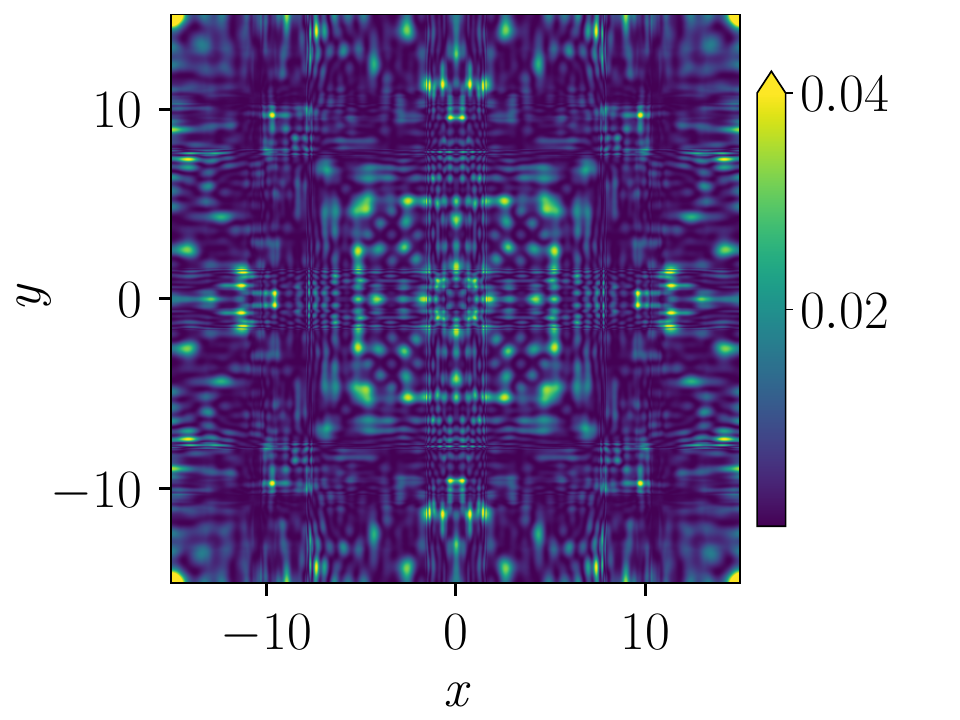}}\\
    \subfloat[$\Z8$, $t = 10$]{\includegraphics[width=0.33\textwidth]{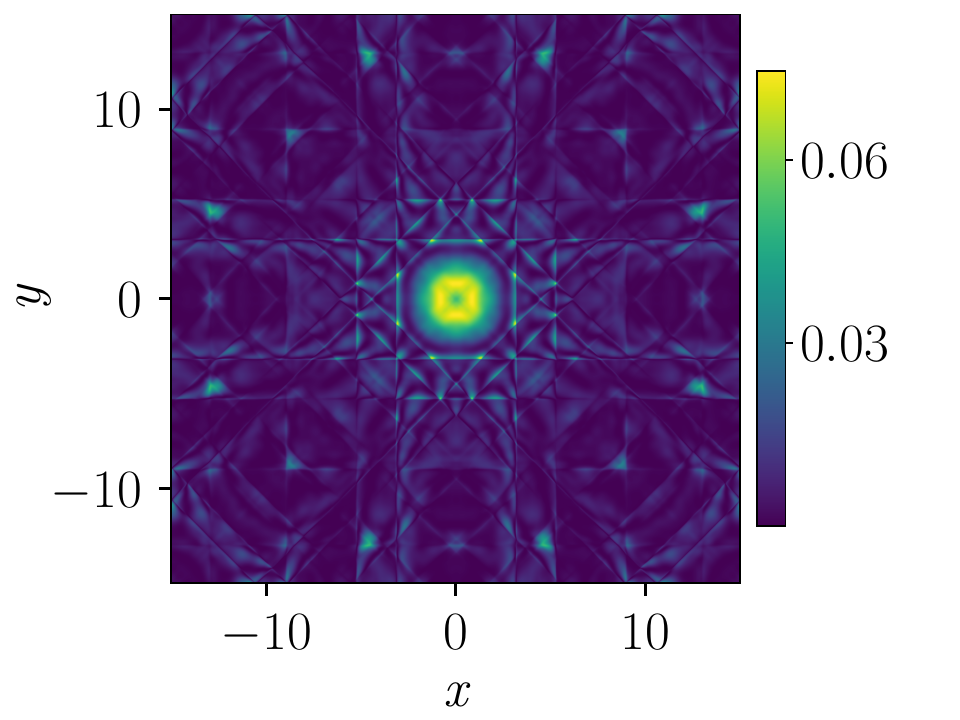}}
    \subfloat[$\Z8$, $t = 20$]{\includegraphics[width=0.33\textwidth]{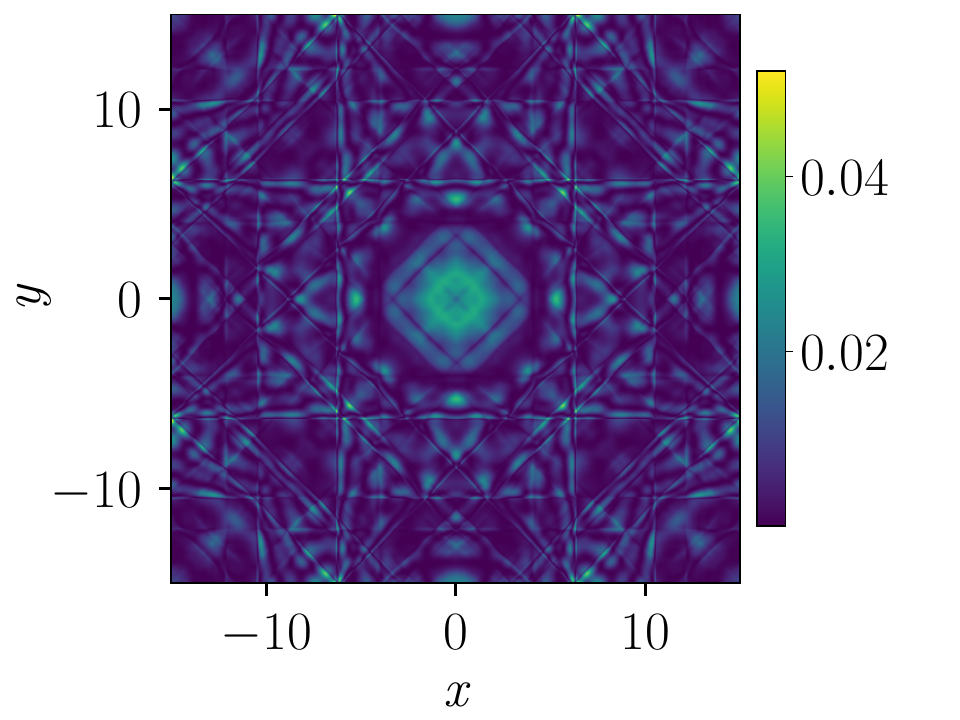}}
    \subfloat[$\Z8$, $t = 100$]{\includegraphics[width=0.33\textwidth]{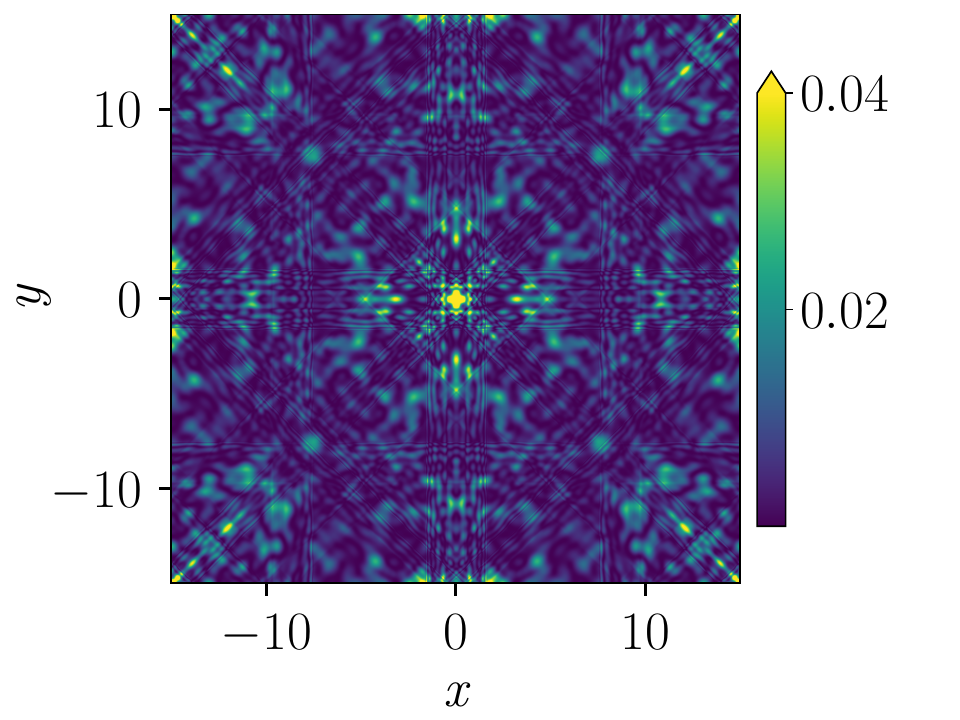}}
    \caption{Dynamics of the $\phi^2$-condensate after a local quench corresponding to the insertion of operator $\phi$. \textit{(a) -- (c)}~Spatial distribution of perturbations for fixed moments of time in the $\Z4$-symmetric fractonic model. \textit{(d) -- (f)}~The same for the $\Z8$-symmetric fractonic model. For all cases, $L_x = L_y = 30$, $m = 1$, $\eps = 0.05$, $\mu_0 = \mu = 1$, $\epsilon = 0.1$.}
    \label{fig:Phi2FinVol}
\end{figure}

The numerical results for the post-quench dynamics are shown in \figref{fig:2PointFinVol}, for the two-point function, and in \figref{fig:Phi2FinVol}, for the $\phi^2$-condensate. Interestingly, the spatial profiles of correlation functions have highly irregular fractal-looking structures, \figref{fig:2PointFinVol}(c)~--~(i). In the case of relativistic field theory, \figref{fig:2PointFinVol}(c), they appear at larger times simply due to self-interference of the propagating wave front after it circumvents the torus several times. However, for fractons it is more than that. Naively, one can expect that it is a similar self-interference effect, because the non-causally propagating excitations instantaneously wind around the torus for infinitely many times and can interfere immediately. At the same time, as one can see from, e.g., \figref{fig:2PointFinVol}(d), the set of equidistant parallel lines cannot be viewed as a periodic continuation of the original ``cross-shaped'' excitation across the boundary. These patterns are intricately connected with mathematical properties of sums like~\eqref{eq:2point_fin_vol} similar to those arising in the description of the Talbot effect in optics~\cite{BerryKlein}. 

\section{Fractional dimension of the wave fronts}
\label{sec:fract_dim}

As clear from Figs.~\ref{fig:2PointFinVol} and~\ref{fig:Phi2FinVol}, evolution of the fractonic model results in rather complicated irregular spatial distributions of the observable values immediately after the quench and, as we discussed above, goes beyond the self-interference effect in contrast with the conventional relativistic model, where highly oscillating patterns emerge at larger times, after the wave front circumvents the torus and interferes with itself. What is interesting to note is that the spatial profiles of $\langle \phi(t,x,y) \phi(0) \rangle$ and $\langle \phi^2(t,x,y) \rangle$ in the fractonic theory are not just highly oscillating, but exhibit multi-scale self-similarity, i.e., can be regarded as fractals. To show that, we resort to studying fractional dimensions of one-dimensional sections of the two-dimensional patterns using the method suggested in~\cite{Dubuc} and previously used by us in~\cite{Ageev:2020-fractal}. To perform the case study, we focus on the evolution of $\langle \phi^2(t,x,y) \rangle$, and for each value of $t$, we take one-dimensional section of the function profile along $x = y$ line. For the sake of completeness, we provide a brief explanation of the method here.

\begin{figure}[!ht]\centering
    \includegraphics[width=0.6\textwidth]{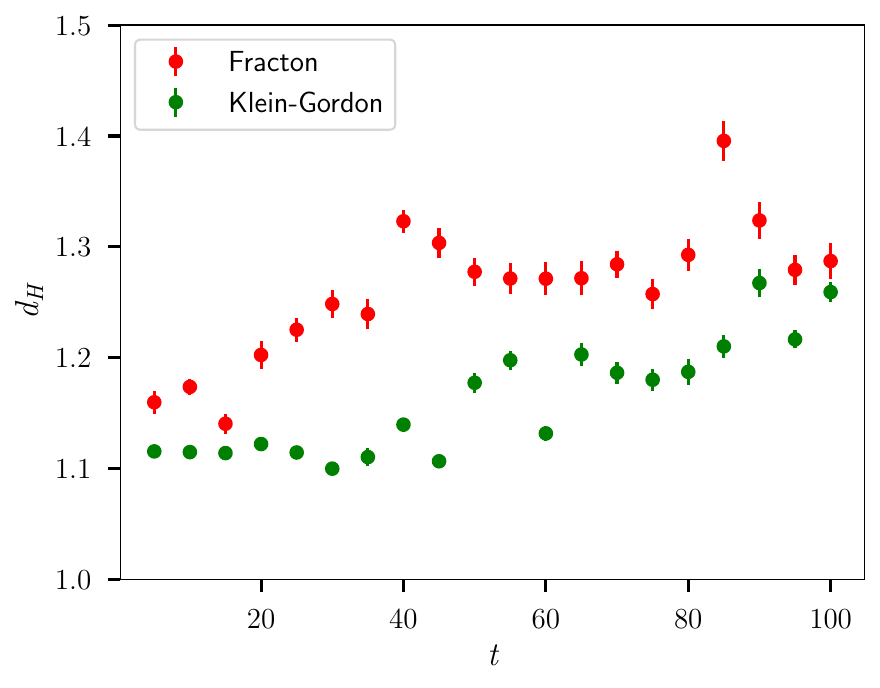}
    \caption{Fractional dimension of the one-dimensional $x = y$ section of $\corrfunc{\phi(t,x,y)^2}$ profile as a function of time for the fractonic and the relativistic theories. The parameters are $L_x = L_y = 30$, $m = 1$, $\eps = 0.02$ for the both cases and $\mu_0 = \mu = 1$, $\epsilon = 0.1$ for the fractonic case.}
    \label{fig:Dubuc}
\end{figure}

To compute fractional dimension of a one-dimensional profile $\Ocal (x) \equiv \corrfunc{\phi^2(t_0, x, x)}$, where coordinate $x$ and value $\Ocal$ have different geometric meanings, the conventional box-counting algorithm cannot be applied. According to~\cite{Dubuc}, the profile needs to be represented as a discrete set of values $\{\Ocal_i\}$ of length~$N$ (in our simulations we discretize it into $N = 5000$ points). Then, for some integer~$\delta$, functions returning the maximal and the minimal values of the set within the~$\delta$ neighborhood of the $i$-th element of the set are defined as
\be
    \begin{aligned}
        & u_\delta(i) = \sup\limits_{i'\in R_\delta (i)} S_{i'}, \\
        & b_\delta(i) = \inf\limits_{i'\in R_\delta (i)} S_{i'},
    \end{aligned}
\ee
where $R_\delta (i) = \{i': |i'-i| \leq \delta, i' \in [1, N]\}$, and periodic boundary conditions on the set of indices are assumed; $u_\delta(i)$ and $b_\delta(i)$ are the upper and the lower enveloping curves of $\{\Ocal_i\}$. Then, an analog of the box-counting function is defined as
\be
    V(\delta) = \frac{1}{\delta^2}\sum\limits_i (u_\delta(i) - b_\delta(i)).
\ee
It returns the number of $\delta \times \delta$ blocks within the band between $u_\delta(i)$ and $b_\delta(i)$ envelops. The fractional dimension $d_H$ is then obtained by fitting this function with a power-law
\be
    V(\delta) = \frac{a}{\delta^{d_H}}.
    \label{eq:fractal_fit}
\ee
It is worth noticing, that the scaling is never perfectly algebraic, but the resulting fit errors are rather small. To estimate $d_H$, we make fit of $V(\delta)$ for a range $\delta \in [2, 80]$. The result for $d_H$ of the one-dimensional sections of $\corrfunc{\phi^2}$ as a function of $t$ is shown in \figref{fig:Dubuc}.

\begin{figure}[!ht]\centering
    \includegraphics[width=0.7\textwidth]{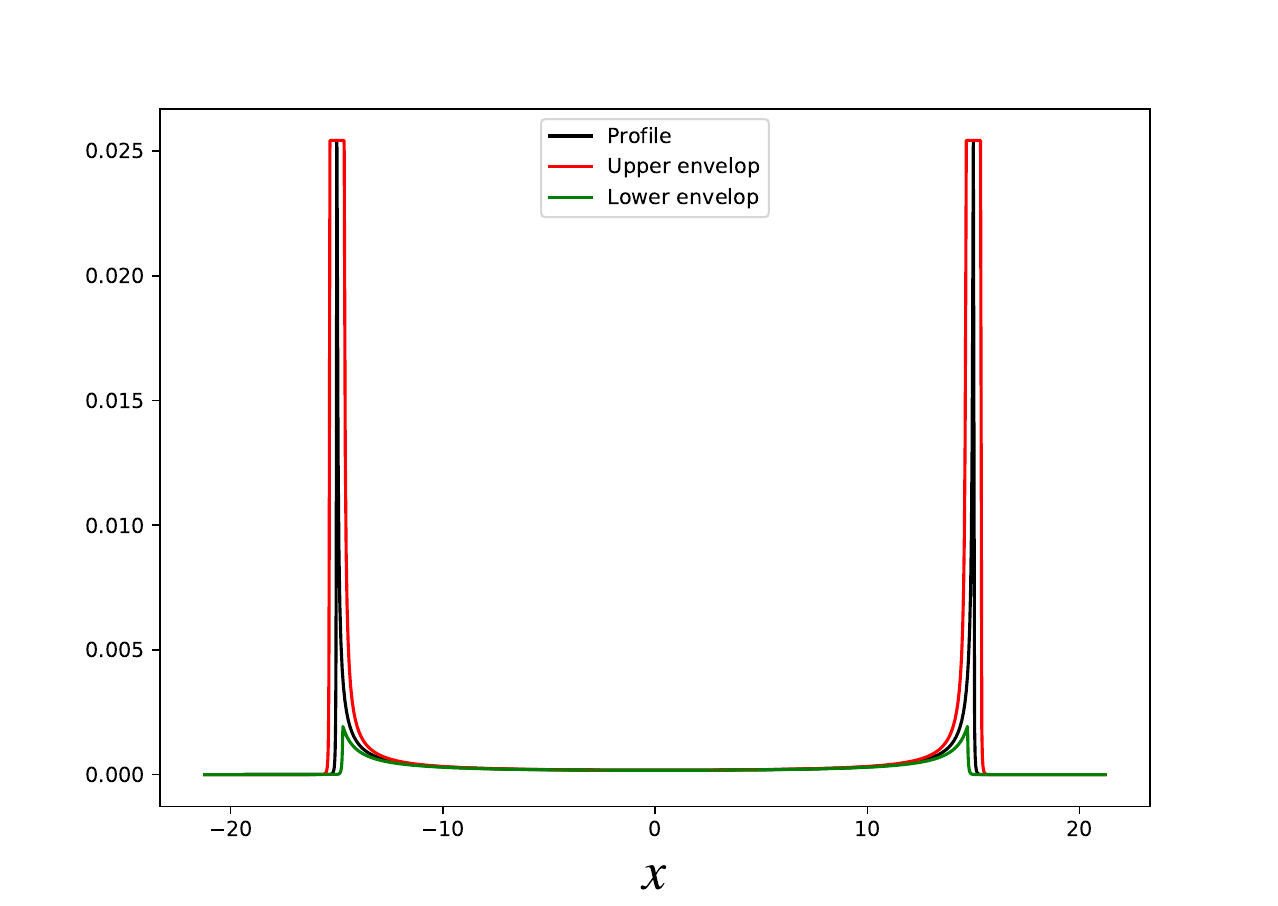}
        \caption{Upper (red) and lower (green) envelops of a smooth, though peaked, curve (black) for $\delta = 40$ (5000 points in total). The non-vanishing value of $V(\delta)$ between the envelops contributes to the fractional dimension estimated by \eqref{eq:fractal_fit}.}
    \label{fig:envelops}
\end{figure}

It should be noted that this method tends to overestimate the fractional dimension and gives $d_H \simeq 1.1$ for regular non-fractal curves with a few sharp peaks as in \figref{fig:envelops} (a typical example of this is the profiles of excitations in the relativistic theory at small times). This artifact is unavoidable when the Hausdorff dimension is estimated numerically. If~$\delta$ value is larger than the characteristic width of the peak, the difference between the upper and the lower envelops is non-negligible, and contributes to~\eqref{eq:fractal_fit} shifting $d_H$ from~1. By taking this into account, we can see from \figref{fig:Dubuc} that the Hausdorff dimension of the wave front in relativistic field theory acquires non-trivial values after a certain period of time, when the signal circumvents the torus and self-interferes. For the fracton field theory, the dimension is rather high from the very beginning, and always exceeds that of the relativistic wave front.

\begin{figure}[!ht]\centering
    \subfloat[local quench]{
    \includegraphics[width=0.45\textwidth]{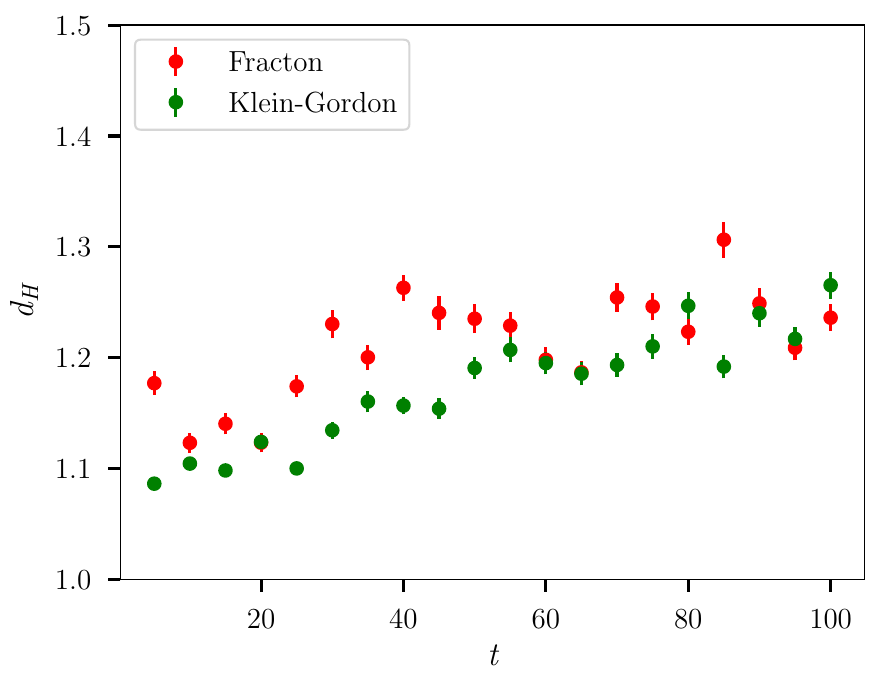}}
    \subfloat[global quench]{\includegraphics[width=0.45\textwidth]{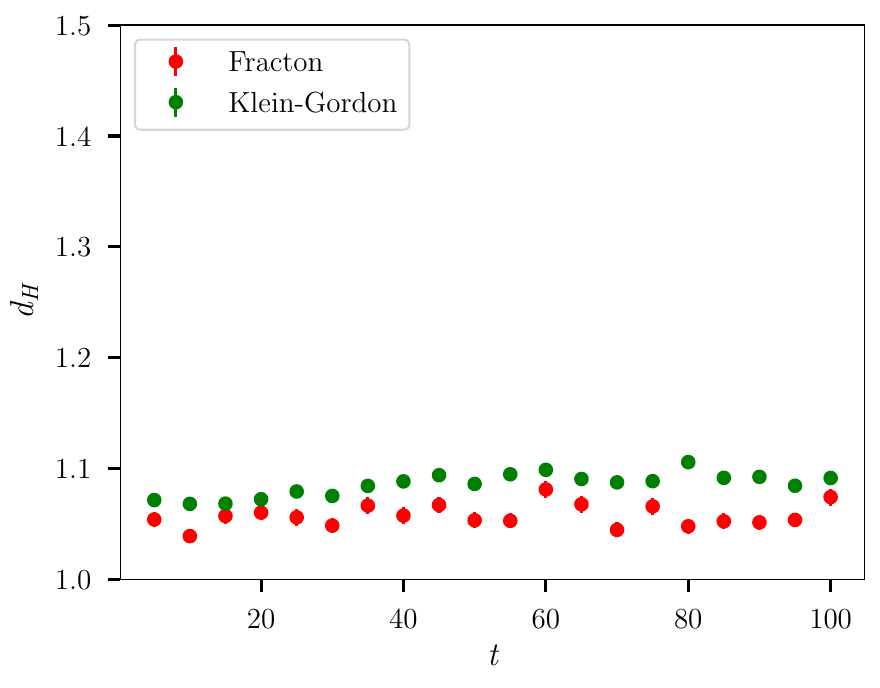}}
    \caption{Fractional dimension of the one-dimensional $x = y$ section of the two-point function profile as a function of time for the fractonic and the relativistic theories in the cases of the local (a) and global (b) quench. The parameters are $L_x = L_y = 30$, $m = 1$, $\eps = 0.02$ (a) and slab width $\tau_0 = 0.01$ (b) for both the relativistic and fractonic cases and $\mu_0 = \mu = 1$, $\epsilon = 0.1$ for the fractonic case.}
    \label{fig:fract_dim_compare}
\end{figure}

Remarkably, it is the analysis of fractional dimensions that allows us to make a comparison between the local quench setup of this paper and the simpler setup of global quench in fractonic theories, which was investigated in~\cite{Ageev:2023wrb}. In that setup, the system is taken out of equilibrium by abrupt change of the parameters or the symmetries of the system. Since the excitation is global, one-point correlation functions depend trivially on spatial coordinates. Hence, we should compare two-point function instead of $\langle \phi^2(t,x,y) \rangle$ correlator. In \figref{fig:fract_dim_compare}, we plotted the corresponding dependencies. In the local quench setup, the fractonic theory fractional dimension still grows faster than in the relativistic theory although the difference is now less pronounced. The global quench setup (here, the boundary global quench, which is characterized by the slab width) comes as complete opposite: not only do fractional dimensions both for the fractonic and the relativistic cases hardly grow but even the dimension in the fractonic case lies lower than for the relativistic one.

This hints to us that even if the dynamics of two systems look quite similar in their irregular character, the systems may demonstrate different behavior of the features characterizing this irregularity.\footnote{For more on this issue, refer to an upcoming paper by D.~Ageev and V.~Pushkarev.}

\section{Discussion and conclusions}
\label{sec:conclusions}

We have analyzed the out-of-equilibrium dynamics following a local point-like quench in fracton field theories, and, rather expected, it turns out to be very distinct from that of conventional relativistic quantum fields. While in the relativistic settings, the initially localized excitation propagates outwards continuously delocalizing, in free fractonic theory the very notion of local quench is somewhat questionable: due to the strong UV/IR mixing, a point-like perturbation instantaneously excites the system along the lines of discrete symmetry of the theory, creating a non-local pattern of correlations in no time. This observation is supported by the fact that the Lieb-Robinson bound for fractonic theories degenerates along the mobility directions in the continuum limit, and no longer imposes causality constraints. By introducing mass and the relativistic regularization term, it is possible to induce causal dynamics of wave fronts in the fractonic theory, however, the non-causal instantaneous propagation of signals does not go away, and these two coexist leading to non-trivial interference patterns. It should be noted that mass and the relativistic regularizaion violate the dipole momentum conservation law --- the defining feature of fractons --- but still allow to obtain a tractable approximate story of what happens during the strongly singular and UV/IR mixed dynamics of the non-regularized theory. In particular, the post-quench dynamics in fracton field theories preserves the $\Z{n}$ symmetry of the theory, as can be seen from the geometry of causal wave fronts. When the free fracton field is placed in a finite volume, fractal-like patterns of excitations emerge across the whole space immediately after the quench. This effect is observed only for locally introduced perturbation, not for the global (as in~\cite{Ageev:2023wrb}) one, and goes beyond self-interference of non-causally propagating signals requiring further investigations.

A few questions arise from these considerations. First of all, while the fractonic theories by definition violate the Lorentz symmetry, the fact that they do it in a way that results in non-causal signal propagation is undesirable. To find a modification of the fractonic model that has a strict propagation speed limit is an important problem. Studying out-of-equilibrium dynamics in an interacting fracton field theory can potentially shed a light on this issue. Along the same line, it is interesting to deeper understand the connection between non-causal propagation and the emergence of fractal excitation patterns. While it is tempting to relate these two aspects, the fact that the interference of instantaneously propagating signal alone cannot explain these structures leads to a question if this effect persists in modifications of fracton field theories with restored causality.

\acknowledgments
The work of D.S.A., A.I.B. and V.V.P. was performed at the Steklov International Mathematical Center and supported by the Ministry of Science and Higher Education of the Russian Federation (agreement no. 075-15-2022-265). D.S.A., A.I.B. and V.V.P. are supported by the Foundation for the Advancement of Theoretical Physics and Mathematics ``BASIS''.

The work of A.A.B. and A.I. was supported by the European Research Council (ERC) under the European Union’s Horizon 2020 research and innovation program, grant agreement 854843-FASTCORR. A.A.B. acknowledges the research program ``Materials for the Quantum Age'' (QuMat) for financial support. This program (registration number 024.005.006) is part of the Gravitation program financed by the Dutch Ministry of Education, Culture and Science (OCW). 

The authors declare that this work has been published as a result of peer-to-peer scientific collaboration between researchers. The provided affiliations represent the actual addresses of the authors in agreement with their digital identifier (ORCID) and cannot be considered as a formal collaboration between the aforementioned institutions.

\appendix

\section{Fractonic propagator on a two-dimensional plane}
\label{app:A}

The action of the Euclidean free massive fracton field theory~\cite{Seiberg:2020bhn, Distler:2021bop} with the relativistic regularization~\cite{Ageev:2023wrb} is given by
\be
    S = \frac{A}{2}\int d\tau\,dx\,dy\left(\mu_0(\partial_{\tau}\phi)^2 + \epsilon(\partial_x\phi)^2 + \epsilon(\partial_y\phi)^2 + \frac{1}{\mu}(\partial_x\partial_y\phi)^2 + \mu m^2\phi^2\right),
    \label{eq:Eaction}
\ee
where $A$ is a normalization constant that we keep here for generality but fix it to be equal to $1/(4\pi)$ in the calculations. The corresponding equation of motion reads
\be
    -\mu_0\ddot\phi - \epsilon\left(\partial^2_x\phi + \partial^2_y\phi\right) + \frac{1}{\mu}\partial^2_x\partial^2_y\phi + \mu m^2\phi = 0.
\ee
In momentum space, it takes the form
\be
   \left(\mu_0\om^2 + \epsilon\left(k^2 + q^2\right) + \frac{k^2q^2}{\mu} + \mu m^2\right)\phi = 0,
    \label{eq:MEOM}
\ee
and the corresponding dispersion relation of the theory is
\be
    \om^2(k,q) = -\frac{1}{\mu_0\mu}\left(\epsilon\mu\left(k^2 + q^2\right) + k^2q^2 + \mu^2 m^2\right).
\ee
The Euclidean two-point function of the massive scalar fractonic field~\eqref{eq:Eaction} is given~by the Fourier transform
\be
    \begin{aligned}
        \corrfunc{\phi(\tau, x, y)\phi(0, 0, 0)} & = \frac{1}{A}\,\mathcal{F}_{\om, k, q}\left[\frac{\mu}{\mu_0\mu\om^2 + \epsilon\mu\left(k^2 + q^2\right) + k^2q^2 + \mu^2 m^2}\right] \equiv \\
        & \equiv \frac{1}{A}\int\frac{d\om\,dk\,dq}{(2\pi)^3}\frac{\mu e^{i\om\tau + ikx + iqy}}{\mu_0\mu\om^2 + \epsilon\mu\left(k^2 + q^2\right) + k^2q^2 + \mu^2 m^2}.
    \end{aligned}
    \label{eq:prop_massive}
\ee
The Fourier transformation with respect to $\om$ gives
\be
    \begin{aligned}
        & \mathcal{F}_{\om}\left[\frac{\mu}{\mu_0\mu\om^2 + \epsilon\mu\left(k^2 + q^2\right) + k^2q^2 + \mu^2 m^2}\right] = \\
        & = \frac{1}{2\mu_0}\sqrt{\frac{\mu_0\mu}{\epsilon\mu\left(k^2 + q^2\right) + k^2q^2 + \mu^2 m^2}} \, \exp\left[-|\tau|\sqrt{\frac{\epsilon\mu\left(k^2 + q^2\right) + k^2q^2 + \mu^2 m^2}{\mu_0\mu}}\right].
    \end{aligned}
\ee
The other two transformations with respect to $k$ and $q$ cannot be calculated analytically, so we leave the correlation function in the mixed representation with explicit dependence on $x$ and $y$ coordinates,
\be
    \begin{aligned}
        & \corrfunc{\phi(\tau, x, y)\phi(0, 0, 0)} = \\
        & = \frac{1}{2A\mu_0}\int\frac{dk\,dq}{(2\pi)^2}\sqrt{\frac{\mu_0\mu}{\epsilon\mu\left(k^2 + q^2\right) + k^2q^2 + \mu^2 m^2}} \, e^{-\sqrt{\tau^2}\sqrt{\frac{\epsilon\mu\left(k^2 + q^2\right) + k^2q^2 + \mu^2 m^2}{\mu_0\mu}} + ikx + iqy}.
    \end{aligned}
    \label{eq:mixedprop}
\ee
Note that we substituted $|\tau|$ with $\sqrt{\tau^2}$ in the final expression in order to make clear how to perform analytical continuation to the Lorentzian time, $\tau \to it$. For a detailed discussion of this aspect, we refer the reader to Apps.~A and B of~\cite{Ageev:2022kpm}.

The integral~\eqref{eq:mixedprop} is divergent for $m = 0$ and $\epsilon = 0$. However, its derivatives are well-defined and can be expressed analytically. After performing integration over $k$ and $q$ in turn, we obtain, in Euclidean signature,
\be
    \partial_{\tau}\corrfunc{\phi(\tau, x, y)\phi(0, 0, 0)} = \frac{\sqrt{\mu \mu_0}}{A (2\pi)^2\tau}\Big(\exp(u)Ei(-u) + \exp(-u)Ei(u)\Big),
\ee
where $u = xy\sqrt{\mu\mu_0}/\tau$, and $Ei(u)$ is the exponential integral. By analytical continuation, we obtain, in Lorentzian signature,
\be
    \partial_{t}\corrfunc{\phi(t, x, y)\phi(0, 0, 0)} = \frac{\sqrt{\mu \mu_0}}{2A\pi^2t}\Big(\cos(u)Ci(u) - \sin(u)Si(u)\Big),
\ee
where $u = xy\sqrt{\mu\mu_0}/t$, and $Si(u)$, $Ci(u)$ are the trigonometric integrals.

\section{Fractonic propagator in the finite-volume theory}
\label{app:B}

Let us consider Euclidean theory~\eqref{eq:Eaction} on a $2 + 1$-dimensional torus $x \sim x + L_x$, $y \sim y + L_y$. Its Green function is a solution to the following system with periodic boundary conditions
\be
    \left\{
    \begin{aligned}
        & A \left( - \mu_0\partial^2_\tau - \epsilon(\partial^2_x + \partial^2_y) - \frac{1}{\mu}\partial^2_x\partial^2_y + \mu m^2 \right) K(\Vec{x}_1 - \Vec{x}_2) = \delta^{(3)}\left(\Vec{x}_1 - \Vec{x}_2\right), \\
        & K\left(\tau, x + L_x, y\right) = K(\tau, x, y), \\
        & K\left(\tau, x, y + L_y\right) = K(\tau, x, y),
    \end{aligned}
    \right.
\ee
where $\Vec{x}_i = \{\tau_i, x_i, y_i\}$ are Euclidean 3-vectors. The solution can be written as an infinite series over the discrete spectrum of modes
\be
    \corrfunc{\phi(\tau, x, y)\phi(0, 0, 0)} = \frac{1}{A L_x L_y}\sum\limits_{n = -\infty}^\infty\sum\limits_{s = -\infty}^\infty\int\frac{d\om}{2\pi}\frac{e^{i\om\tau + i k_n x + i q_s y}}{\mu_0\om^2 + \epsilon\left(k_n^2 + q_s^2\right) + \frac{1}{\mu}k_n^2 q_s^2 + \mu m^2},
\ee
where $k_n = 2\pi n / L_x$ and $q_s = 2\pi s / L_y$ are Matsubara-like frequencies. It is only possible to evaluate analytically the integral over~$\om$
\be
    \int\frac{d\om}{2\pi}\,\frac{e^{i\om\tau}}{\om^2 + \om_{ns}^2} = \frac{1}{2|\om_{ns}|}e^{-|\om_{ns}| \cdot |\tau|} \equiv \frac{1}{2|\om_{ns}|}e^{-|\om_{ns}|\sqrt{\tau^2}},
    \label{eq:how_to_continue}
\ee
which results in
\be
    \begin{aligned}
        & \corrfunc{\phi(\tau, x, y)\phi(0, 0, 0)} = \\
        & = \frac{1}{2A L_x L_y \mu_0}\sum\limits_{n = -\infty}^\infty\sum\limits_{s = -\infty}^\infty\sqrt{\frac{\mu_0\mu}{\epsilon\mu\left(k_n^2 + q_s^2\right) + k_n^2q_s^2 + \mu^2m^2}}e^{-\sqrt{\tau^2}\sqrt{\frac{\epsilon\mu\left(k_n^2 + q_s^2\right) + k_n^2q_s^2 + \mu^2m^2}{\mu_0\mu}} + i k_n x + i q_s y},
    \end{aligned}
    \label{eq:fracton_2point_fin_vol}
\ee
where we substitute $|\tau| \to \sqrt{\tau^2}$ as in~\eqref{eq:mixedprop}.

In fact, \eqref{eq:fracton_2point_fin_vol} can be generalized onto the case of arbitrary free field dispersion relation $\om_{ns} \equiv \om(k_n, q_s)$,
\be
    \corrfunc{\phi(\tau, x, y)\phi(0, 0, 0)} = \frac{1}{A L_x L_y}\sum\limits_{n = -\infty}^\infty\sum\limits_{s = -\infty}^\infty\frac{e^{-\om_{ns}\sqrt{\tau^2} + i k_n x + i q_s y}}{2\om_{ns}}\Bigg|_{k_n = \frac{2\pi n}{L_x}, \,\, q_s = \frac{2\pi s}{L_y}}.
    \label{eq:2point_fin_vol}
\ee
Taking into account that both relativistic, $\om_{ns} = \sqrt{k_n^2 + q_s^2 + m^2}$, and regularized fractonic, $\om_{ns} = \sqrt{\epsilon\mu(k_n^2 + q^2_s) + k_n^2q_s^2 + \mu^2m^2}/\sqrt{\mu_0\mu}$, discrete dispersion relations are symmetric under the sign-inversion of $n$ and $s$, we can transform this expression to
\be
    \begin{aligned}
        \corrfunc{\phi(\tau, x, y)\phi(0, 0, 0)} & = \frac{1}{A L_x L_y} \cdot \frac{e^{-\om_0\sqrt{\tau^2}}}{2\om_0} + \frac{1}{A L_x L_y}\sum_{n = 1}^{\infty} (\cos(k_n x) + \cos(k_n y))\cdot \frac{e^{-\om_n\sqrt{\tau^2}}}{\om_n} + \\
        & + \frac{2}{A L_x L_y}\sum_{n = 1}^{\infty}\sum_{s = 1}^{\infty} \cos(k_n x)\cos(q_s y) \cdot \frac{e^{-\om_{ns}\sqrt{\tau^2}}}{\om_{ns}},
    \end{aligned}
\ee
where $\om_0 = m$ and $\om_n = \sqrt{k_n^2 + m^2}$ for the relativistic dispersion relation, and $\om_{n} = \sqrt{\epsilon\mu k_n^2 + \mu^2 m^2}/\sqrt{\mu\mu_0}$ for the regularized fractonic dispersion relation.

\section{Fractonic energy density and dipole moment}
\label{app:C}

In the mostly minus signature, the Lagrangian density of the relativistic massive scalar field theory is given by
\be
    \Lcal = \frac{1}{2}g^{\mu\nu}\partial_\mu\phi\partial_\nu\phi - \frac{1}{2}m^2\phi^2 = \frac{1}{2}\left(\dot{\phi}^2 - (\nabla\phi)^2 - m^2\phi^2\right),
\ee
leading to the energy density following from the Noether theorem
\be
    \Ecal = \frac{\delta\Lcal}{\delta(\partial^0 \phi)}\partial_0 \phi - g_{00}\Lcal = \frac{1}{2}\left(\dot{\phi}^2 + (\nabla\phi)^2 + m^2\phi^2\right).
\ee

In the mostly plus signature, the corresponding Lagrangian density and the energy density take the form
\be
    \Lcal = -\frac{1}{2}g^{\mu\nu}\partial_\mu\phi\partial_\nu\phi - \frac{1}{2}m^2\phi^2 = \frac{1}{2}\left(\dot{\phi}^2 - (\nabla\phi)^2 - m^2\phi^2\right),
\ee
and
\be
    \Ecal = -\frac{\delta\Lcal}{\delta(\partial^0 \phi)}\partial_0\phi + g_{00}\Lcal = \frac{1}{2}\left(\dot{\phi}^2 + (\nabla\phi)^2 + m^2\phi^2\right).
\ee
In the Euclidean metric ($t \to -i\tau$, $g_{00} \to -g_{00}$, $\partial_0\phi \to i\partial_0\phi$), we get
\be
    \Lcal = \frac{1}{2}\left(\dot{\phi}^2 + (\nabla\phi)^2 + m^2\phi^2\right),
\ee
and
\be
    \Ecal = -\frac{\delta\Lcal}{\delta(\partial^0 \phi)}\partial_0\phi + g_{00}\Lcal = \frac{1}{2}\left(-\dot{\phi}^2 + (\nabla\phi)^2 + m^2\phi^2\right).
\ee

\bigskip

In the same way, we obtain the following expressions for the Euclidean fracton field theory,
\be
    \Lcal = \frac{1}{2}\left(\mu_0(\partial_{\tau}\phi)^2 + \frac{1}{\mu}(\partial_x\partial_y\phi)^2 + \mu m^2\phi^2\right),
\ee
and
\be
    \Ecal = -\frac{\delta\Lcal}{\delta(\partial^0 \phi)}\partial_0\phi + g_{00}\Lcal = \frac{1}{2}\left(-\mu_0\dot{\phi}^2 + \frac{1}{\mu}(\partial_x\partial_y\phi)^2 + \mu m^2\phi^2\right).
\ee

\bigskip

Then, the Euclidean energy density can be calculated by applying Wick's theorem to the two-point function,
\be
    \begin{aligned}
        \corrfunc{\Ecal}_\phi & = -\frac{\corrfunc{\phi(\eps, 0, 0)\partial_\tau\phi(\tau, x, y)}\corrfunc{\partial_\tau\phi(\tau, x, y)\phi(-\eps, 0, 0)}}{\corrfunc{\phi(\eps, 0, 0)\phi(-\eps, 0, 0)}} + \\
        & + \frac{\corrfunc{\phi(\eps, 0, 0)\partial_x\phi(\tau, x, y)}\corrfunc{\partial_x\phi(\tau, x, y)\phi(-\eps, 0, 0)}}{\corrfunc{\phi(\eps, 0, 0)\phi(-\eps, 0, 0)}} + \\
        & + \frac{\corrfunc{\phi(\eps, 0, 0)\partial_y\phi(\tau, x, y)}\corrfunc{\partial_y\phi(\tau, x, y)\phi(-\eps, 0, 0)}}{\corrfunc{\phi(\eps, 0, 0)\phi(-\eps, 0, 0)}} + \\
        & + \frac{m^2\corrfunc{\phi(\eps, 0, 0)\phi(\tau, x, y)}\corrfunc{\phi(\tau, x, y)\phi(-\eps, 0, 0)}}{\corrfunc{\phi(\eps, 0, 0)\phi(-\eps, 0, 0)}}.
    \end{aligned}
\ee

The remaining correlation functions are to be calculated from the momentum-space two-point function by Fourier transform,
\be
    \corrfunc{\partial_\tau\phi(\tau, x, y)\phi(\tau_0, 0, 0)} = -\frac{\tau - \tau_0}{2A\sqrt{(\tau - \tau_0)^2}}\int\frac{dk\,dq}{(2\pi)^2}e^{-\om\sqrt{(\tau - \tau_0)^2} + ikx + iqy},
\ee
\be
    \corrfunc{\partial_x\phi(\tau, x, y)\phi(\tau_0, 0, 0)} = \frac{i}{A}\int\frac{dk\,dq}{(2\pi)^2}\frac{k\,e^{-\om\sqrt{(\tau - \tau_0)^2} + ikx + iqy}}{2\om}
\ee
and
\be
    \corrfunc{\partial_x\partial_y\phi(\tau, x, y)\phi(\tau_0, 0, 0)} = -\frac{1}{A}\int\frac{dk\,dq}{(2\pi)^2}\frac{kq\,e^{-\om\sqrt{(\tau - \tau_0)^2} + ikx + iqy}}{2\om}.
\ee

\skipline

We are also interested in the dipole momentum, which is globally conserved in pure fractonic theories. In the Lorentzian signature, the dipole current is given by~\cite{Seiberg:2020bhn}
\be
    \begin{aligned}
        J_0 & = \mu_0\partial_t \phi, \\
        J^{xy} & = -\frac{1}{\mu}\partial^{x}\partial^{y}\phi
    \end{aligned}
\ee
with the conservation law
\be
    \partial_0 J = \partial_x \partial_y J^{xy}.
\ee
In this paper, we consider the square of the zero component of the current, $J^2_0$,
\be
    \corrfunc{J^2_0} = \frac{\bra{\phi(i\eps, 0, 0)} (\partial_t \phi_0(t, x, y))^2 \ket{\phi(-i\eps, 0, 0)}}{\corrfunc{\phi(i\eps, 0, 0)\phi(-i\eps, 0, 0)}}.
\ee
The correlation function has a form similar to that of $\phi^2$-condensate~\eqref{eq:phi2_corr_def}, and thus can be calculated in the same way,
\be
    \corrfunc{J^2_0} = \frac{\left|\int dk\,dq\,\,e^{-\om\sqrt{(\eps - it)^2} + ikx + iqy}\right|^2}{4A\pi^2\int dk\,dq \,\, \om^{-1}e^{-2\eps\om}}.
\ee
In the case of finite volume, the integral over momenta is changed to the sum over Fourier modes.


\bibliography{main}
\bibliographystyle{JHEP}

\end{document}